\documentclass[preprint]{aastex}
\newcommand{\mybibstyle}{apj}
\newcommand{\mytab}{\begin{table}[htb]}
\newcommand{\myfig}{\begin{figure}[htbp]}

\usepackage{amsmath}
\usepackage[section] {placeins}
\usepackage{multirow}

\newcommand{\msun}{\mathrm{M}_\odot}
\newcommand{\bbbar}{$\mathit{b\bar{b}}\;$}
\newcommand{\mumu}{$\mathit{\mu^+\mu^-}\;$}
\graphicspath{{show/}}

\begin{document}
\title{Evidence for extended gamma-ray emission from galaxy clusters}

\author{Jiaxin Han\altaffilmark{1,2,3}, Carlos S. Frenk\altaffilmark{3}, 
Vincent R. Eke\altaffilmark{3}, Liang Gao\altaffilmark{4,3} and Simon D. M. White\altaffilmark{5}}
\email{jxhan@shao.ac.cn}

\altaffiltext{1}{Key Laboratory for Research in Galaxies and Cosmology,
Shanghai Astronomical Observatory, Shanghai 200030, China}
\altaffiltext{2}{Graduate School of the Chinese Academy of Sciences,
19A, Yuquan Road, Beijing, China}
\altaffiltext{3}{Institute of Computational Cosmology, Department of Physics,
University of Durham, Science Laboratories, South Road, Durham DH1
3LE}
\altaffiltext{4}{Partner Group of the Max Planck Institute for Astrophysics,
National Astronomical Observatories, Chinese Academy of Sciences,
Beijing, 100012, China}
\altaffiltext{5}{Max-Planck Institute for Astrophysics,
 Karl-Schwarzschild Str. 1, D-85748, Garching, Germany}
 
\begin{abstract}\noindent We report evidence for extended gamma-ray emission from the Virgo,
Fornax and Coma clusters based on a maximum-likelihood analysis of the
3-year Fermi-LAT data. For all three clusters, excess emission is
observed within three degrees of the center, peaking at the GeV
scale. This emission cannot be accounted for by known Fermi sources or
by the galactic and extragalactic backgrounds. If interpreted as
annihilation emission from supersymmetric dark matter (DM) particles,
the data prefer models with a particle mass in the range $20-60$~GeV
annihilating into the \bbbar channel, or $2-10$~GeV and $>1$~TeV
annihilating into \mumu final states. Our results are consistent with
those obtained by Hooper and Linden from a recent analysis of
Fermi-LAT data in the region of the Galactic Centre. An extended DM
annihilation profile dominated by emission from substructures is
preferred over a simple point source model. The significance of DM
detection is $4.4\sigma$ in Virgo and lower in the other two
clusters. We also consider the possibility that the excess emission
arises from cosmic ray (CR) induced gamma-rays, and infer a CR level
within a factor of three of that expected from analytical
models. However, the significance of a CR component is lower than the
significance of a DM component, and there is no need for such a CR
component in the presence of a DM component in the preferred DM mass
range. We also set flux and cross-section upper limits for DM
annihilation into the \bbbar and \mumu channels in all three clusters.\end{abstract}


\maketitle

\section{Introduction}
The existence of dark matter (DM) in the universe has so far only been
deduced and constrained from its gravitational effect, due to the lack
of electromagnetic interactions of the DM with itself or with baryonic
matter. There are several elementary particle candidates for DM in
various extension of the standard model of particle physics. Weakly
interacting massive particles (WIMPs) are one of the most promising
classes of dark matter candidates, with a self-interaction
cross-section estimated at around $3\times 10^{-26} \rm{cm^3 s^{-1}}$ under
simplified assumptions. These particles arise naturally in theories
that seek to extend the standard model and, at the same time, produce
the correct relic density of dark matter in the early universe. Within
the framework of the minimal supersymmetric standard model (MSSM), the
lightest neutralino emerges as the prototype of a WIMP that is stable
over cosmological timescales and can annihilate into standard model
particles. These particles are generally known as cold dark matter.

Much effort has been devoted to the search for dark matter, either
directly or indirectly. Direct detection involves identifying the rare
events of DM scattering off ordinary matter or searching for new
particles near the weak scale at the LHC. Indirect detection involves
looking for the annihilation or decay products of dark matter in
cosmic rays and gamma rays. In particular, pair annihilation produces
gamma-ray photons at a rate proportional to the square of the dark
matter density, which then propagate, almost without absorption, to the
observer. In this case, the Galactic Center should be the brightest
gamma-ray source on the sky \citep[][and references
therein]{Volker}. There have been recent claims of extended emission
(besides a central point source) in the Galactic Center
\citep{GalCen,GalCen1,GalCen2,GalCen3}, which is consistent with
either DM annihilation from a halo with a slightly cuspier density
profile (an inner slope of $\sim -1.25$ to $-1.4$) than the typical NFW
cold dark matter galactic halo \citep{NFW96,NFW97}, or with proton collisions
accelerated by the central galactic black hole. The latter process is
very poorly understood so a definitive conclusion is not yet
possible. 

Targetting the entire sky rather than the Galactic center in searching
for annihilation radiation may seem a good strategy since it takes
advantage of the large-scale distribution of dark matter while
avoiding some of the uncertainties arising from the astrophysical
modelling of galactic gamma-ray sources. However, the fact that we are
located near the center of the Galactic halo and that most of the
annihilation emission outside the Galactic Center is produced by dark
matter substructures
\citep{Volker,Diemand07} results in a gamma-ray map from annihilation that is almost
uniform on large scales. This makes detection within the Milky Way
halo a difficult task, exacerbated by the additional uncertainty in
having to model the extragalactic background, which is more important
on large scales~\citep{FermiLarge,robust}.

Dwarf galaxies are the most DM-dominated objects known, are
relatively free from astrophysical contamination and appear compact on
the sky. They are therefore promising targets for DM annihilation 
detection. Recent joint analyses of eight to ten dwarf galaxies
\citep{Dwarf,DwarfFermi} resulted in a null detection but have  began
to rule out the canonical annihilation cross-section of $3\times
10^{-26}\rm{cm}^3\rm{s}^{-1}$ for DM masses below $\sim20$~GeV.

Galaxy clusters are the most massive virialized DM structures in the
universe and are also good targets for indirect DM searches. The
presence of a large population of DM substructures (or subhalos)
predicted by numerical simulations further boosts the detectability of
DM in clusters. Although the total mass within subhalos amounts to
only 10 to 20 percent of the total halo mass, the density enhancement
within subhalos can boost the total cluster annihilation luminosity by
a factor as high as 1000 when extrapolated down to a subhalo mass
limit of one Earth mass as expected, for example, for a $\sim 100$~GeV WIMP
\citep{Gao,Pinzke2}. As the distribution of subhalos is much less
concentrated than that of the smooth main halo, the total annihilation
emission from clusters is predicted to be extended. Thus, attempts to
detect DM annihilation assuming a point source or NFW-squared type
profile could miss most of the signal. In fact, just such a search
using the 11-month Fermi-LAT data has yielded a null detection in six
clusters \citep{FermiCluster}

In this work, we use the 3-year Fermi-LAT data to search for extended
emission from clusters. We consider possible
contributions from cosmic ray (CR) induced gamma-ray emission, which
can be as high as, or higher than, that from cluster DM annihilation
\citep{CRvsDM,Pinzke1,Pinzke2}. We adopt the model proposed by
\citet{Gao} for the cluster annihilation 
profile and the semi-analytic method developed by \citet{Pinzke1} to
model CR induced gamma-ray emission and provide constraints on both
the DM and CR components. We focus on three galaxy clusters: Coma,
which is predicted to have the highest signal-to-noise according to
\citet{Gao}, and Fornax and Virgo which are predicted to have the
lowest astrophysical contamination \citep{Pinzke2}.

%Marginally significant detection is
%found in Virgo and evidence of excess extended emission in both
%spatial and energy space are observed in other two clusters as well.

During the final stages of preparation of this work, a paper 
\citep{Huang} was posted on arxiv presenting 
a null detection of DM annihilation emission from a combined analysis
of eight galaxy clusters. Our work differs from this interesting paper
in several respects: firstly, we assume a DM annihilation profile
based on high resolution cosmological simulations \citep{Gao};
secondly, we assess the impact of cosmic rays in the detection of dark
matter; and finally, we include the Virgo cluster in our sample which
turns out to be the best candidate. The constraints we set on the
annihilation cross-section are consistent with those in \citet{Huang}.
 
The cosmological parameters used in this work are the same as those
assumed by \citet{Gao}:
$\Omega_m=0.25$,$\Omega_\Lambda=0.75$,$h=0.73$.

\section{Modeling gamma-ray emission in clusters }\label{sec_model}

\myfig
\includegraphics[scale=0.3]{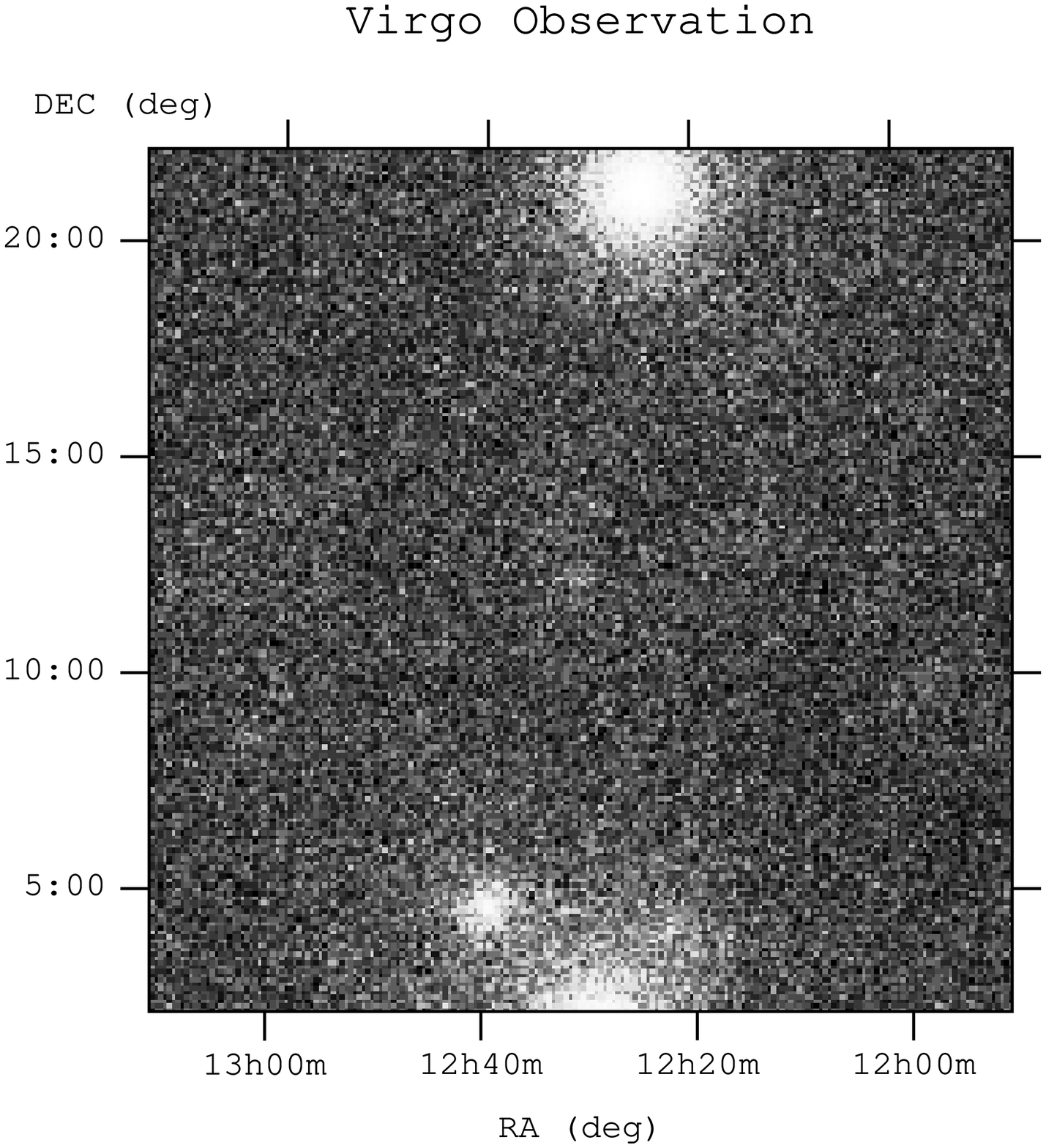}%
\includegraphics[scale=0.5]{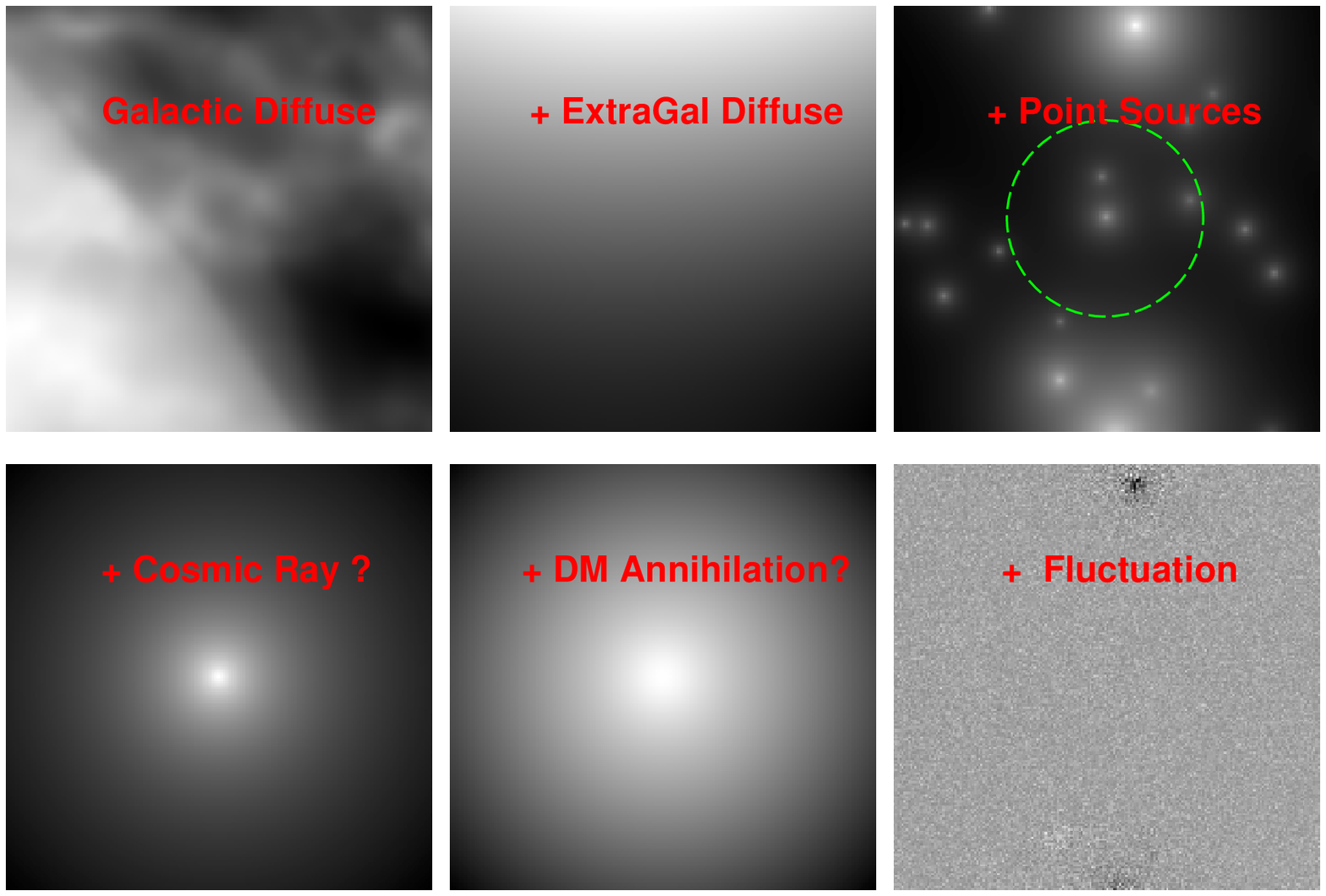}
\caption{Decomposition of the Fermi-LAT image in the region of the
Virgo cluster into model components. The observed photon count image
from 100MeV to 100~GeV is shown on the left. The right panels show the
integrated image over the same energy range for the various model
components: galactic diffuse emission, extragalactic diffuse emission,
2FGL point sources, cosmic-ray photons and DM annihilation emission,
as labeled. The green circle in the ``Point Sources'' panel marks the
virial radius of the cluster. The ``Fluctuation" panel shows the
residual image for our best-fit DM model. The images have been
enhanced individually in color space for contrast.}\label{f_components}
\end{figure}

We model the observed gamma-ray emission in clusters with several
components as shown in Figure~\ref{f_components}: the
galactic foreground (GAL), the extragalactic background (EG),
emission from known point sources,  DM annihilation and CR-induced
emission. The GAL and EG diffuse emission are given by the most recent
templates, \texttt{gal\_2yearp7v6\_v0.fits} and
\texttt{iso\_p7v6source.txt}, which can be obtained from the Fermi-LAT
data server, while the point sources are taken from the LAT 2-year point
source catalogue (2FGL,\citet{2FGL}). We now describe in detail our
models for DM annihilation and CR emission.

\subsection{Cluster annihilation emission}

The gamma-ray intensity along the line-of-sight due to DM annihilation
is given by: 

\begin{equation}\label{eq_I}
I=\frac{1}{8\pi}\sum_f\frac{dN_f}{dE}\sigma_f v\int_{l.o.s.}
(\frac{\rho_\chi}{M_\chi})^2(l)dl, 
\end{equation}
where $M_\chi$ is the DM particle mass, $\displaystyle\frac{dN_f}{dE}$
the particle model dependent term giving the differential number of
photons produced from each annihilation event as a function of energy,
$E$, in a particular annihilation channel, $f$, and $\sigma_f v$ is the
cross-section (or annihilation rate) for that channel, which is
predicted to be constant for cold dark matter. The line-of-sight
integration of the density squared is often expressed in terms of a
dimensionless factor, 
\begin{equation}\label{eq_J}
J=\frac{1}{8.5\rm{kpc}}(\frac{1}{0.3\rm{GeV}/\rm{cm}^3})^2\int_{l.o.s.}
\rho_\chi^2(l)dl. 
\end{equation}
If the source size is much smaller than the instrumental beam size, a
point source approximation is applicable. In this case, the
integration of $J$ over a large enough solid angle is used to
determine the total flux for the point source,
$\mathcal{J}_{int}=\int_{\Delta \Omega} J d\Omega$.

The cluster annihilation emission is modeled with both a point source
approximation and with the extended profile suggested by
\citet{Gao}. We will refer to models using these two profiles
respectively as PT and EXT. If the cluster follows a smooth NFW
profile, then its integrated $J$ factor which determines the total
annihilation flux can be found as
\begin{equation}\label{eq_JNFW}
\mathcal{J}_{NFW}=\frac{4\pi}{3} \rho_s^2 r_s^3\frac{1}{D_A^2} \times
\frac{1}{8.5\rm{kpc}}(\frac{1}{0.3\rm{GeV}/\rm{cm}^3})^2. 
\end{equation}
Here $D_A$ is the angular diameter distance of the cluster and
$\rho_s$ and $r_s$ are the characteristic density and radius for the NFW
profile. They are related to halo concentration and virial radius
through the relations, 
$\rho_s=\dfrac{200}{3}\dfrac{c^3\rho_c}{{\rm log}(1+c)-c/(1+c)}$ and
$r_s=r_{200}/c$, with $\rho_c$ the critical density of the universe,
$r_{200}$ the cluster virial radius within which the average density
is $200\rho_c$ and the concentration parameter, $c$, given by the
following mass-concentration relation: 
\begin{equation}
c=5.74(\frac{M_{200}}{2\times 10^{12}h^{-1}\msun})^{-0.097}  
\end{equation}
\citep{M-c}. Extrapolating to a cutoff mass of $10^{-6}\msun$, the existence of
subhalos will boost this flux by a factor 
\begin{equation}
b(M_{200})=\mathcal{J}_{sub}/\mathcal{J}_{NFW}=1.6\times
10^{-3}(M_{200}/\msun)^{0.39} 
\end{equation}
\cite{Gao}. Using the results of the simulations by these authors, 
the surface brightness profile of subhalo emission can be
fitted within $r_{200}$ by the following formula: 
\begin{equation}
J_{sub}(r)=\frac{16b(M_{200})\mathcal{J}_{NFW}}{\pi
\ln(17)}\frac{1}{r_{200}^2+16r^2}\;\;\;\;\; (r \leq r_{200}). 
\end{equation}
Below we fit the subhalo emission surface brightness beyond the
virial radius and extrapolate to several times the virial radius using
an exponential decay, 
\begin{equation}
J_{sub}(r)=J_{sub}(r_{200})e^{-2.377(r/r_{200}-1)}\;\;\;\;\; (r \geq
r_{200}). 
\end{equation}
The total annihilation profile is the sum of the contributions from a
smooth NFW profile and the subhalo emission. This is completely
dominated by subhalo emission except in the very center of the
cluster. We show the total annihilation profile and its decomposition
into main halo and subhalo contributions in the left panel of
Figure~\ref{fig_profile}, taking Virgo as an example. This profile is
further inflated after convolution with the LAT point spread function.

Following \citet{FermiCluster}, we consider two representative
annihilation channels, namely into $b-\bar{b}$ and
$\mu^+-\mu^-$ final states. The annihilation spectrum is calculated
using the DarkSUSY 
package \citep{DarkSUSY},\footnote{http://www.darksusy.org.} which
tabulates simulation results from 
PYTHIA.\footnote{http://home.thep.lu.se/~torbjorn/Pythia.html} We also
include the contribution from inverse Compton (IC) scattered photons
by energetic electron-positron pairs produced during the annihilation
process, following the procedure described in \citet{Pinzke2}. In
general three external energy sources are involved in the dissipation
and scattering of the injected electrons from annihilation: the Cosmic
Microwave Background (CMB), infrared to UV light from stars and dust,
and the interstellar magnetic field. However, as shown in
\citet{Pinzke2}, the latter two components are expected to be
important only in the inner region of clusters ($<0.03R_{200c}$),
corresponding to less than 0.1 degrees for our three
clusters. Including them would introduce a position-dependent
component to the annihilation spectrum, so for simplicity we only
consider the contribution of CMB photons in the IC calculation. For
the \bbbar channel, IC photons only contribute significantly to the
low energy spectrum for relatively high neutralino mass, while for the
\mumu channel, which has plenty of energetic electrons, the IC emission
can completely dominate the annihilation emission over the full
energy range of interest for the highest neutralino masses 
considered. 

We note that the electroweak corrections recently proposed by
\citet{PPPC_theo} (see also \citet{PPPC_fit}) can bring visible
differences to the \mumu channel spectrum at high WIMP masses before IC
scattering. However, since IC photons dominate at the high mass end
and the electroweak correction only significantly changes the positron
yields at low energy, thus having little effect on the IC spectrum,
the electroweak correction to the total spectrum is still
negligible. The total photon yields are shown in
Figure~\ref{fig_yield}. The almost flat spectrum with a cutoff around
the energy corresponding to the WIMP mass comes from prompt
annihilation emission including continuum secondary photons and final
state radiation from charged final state particles. The low energy rise
originates from IC scattered CMB photons. 

\myfig
\includegraphics[width=0.5\textwidth]{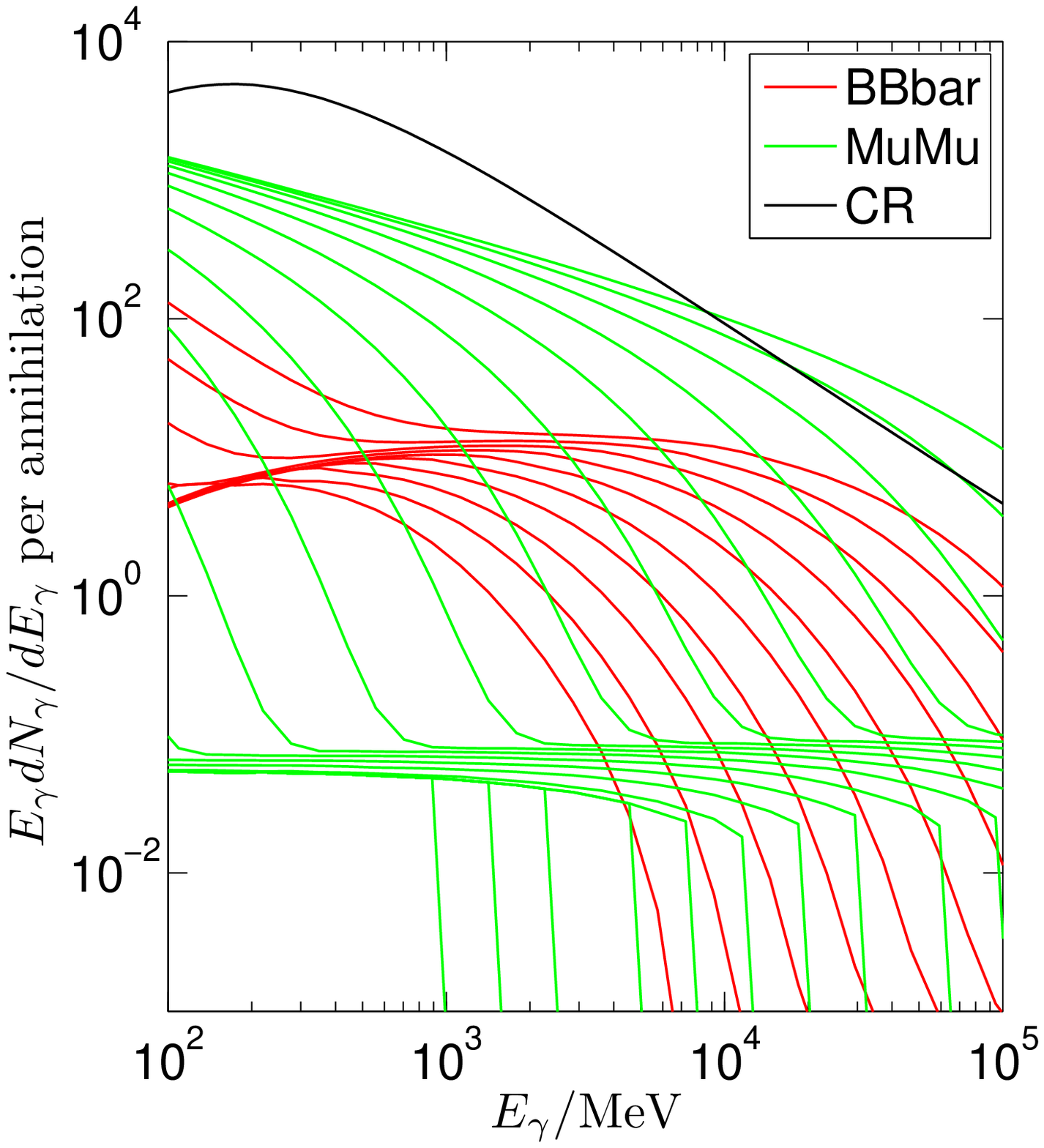}
\caption{Photon yields for \bbbar (red lines) and \mumu (green lines)
channels. Plotted are the total photon yields including continuum
secondary photons, final state radiation from charged final state
particles, as well as inverse Compton scattering of CMB photons by
electron/positron pairs, for the mass range $10-1000$~GeV\ for the
\bbbar channel and $1\rm{GeV}-10\rm{TeV}$ for the \mumu channel. The
masses are sampled uniformly in a log scale. Note that each spectrum
cuts off at an energy corresponding to the particle mass. For
comparison, the black line shows the photon spectrum from cosmic ray
induced photons with arbitrary normalization.}\label{fig_yield}
\end{figure}

\subsection{Cosmic-ray induced gamma-ray emission within clusters}

The cosmic ray induced gamma-ray emission is calculated following
a semi-analytic prescription, derived from high resolution numerical
simulations of galaxy clusters, that models cosmic ray physics self
consistently \citep{Pinzke1}. The gamma-ray photon production rate (or source function)
from pion decay is found to be separable into a
spatial and a spectral part: 

\begin{equation}
q_{CR}(r,E)\equiv\frac{dN_{\gamma}}{dtdVdE}=A(r)s(E), 
\end{equation}
where the spatial part, $A(r)$, is proportional to the square of the
gas density profile multiplied by a slowly varying radial function
parametrized by cluster mass. The spectral part, $s(E)$, is almost
independent of cluster mass and has a power-law form, 
$dN_{\gamma}/d\ln(E_{\gamma})\propto E_{\gamma}^{-1.3}$, for the energy
range $1\sim100$~GeV but flattens at low energies, as shown in
Figure~\ref{fig_yield}. We summarize the detailed form of $A(r)$ and
$s(E)$ plus the gas density profile for the three clusters derived
from X-ray observations in the Appendix. 

The differential gamma-ray flux from this source function is simply
its integral along the line-of-sight:

\begin{equation}
I_{CR}(r,E)=\int_{l.o.s} \frac{q_{CR}(r,E)}{4\pi} dl. 
\end{equation}
This prescription is derived from the average emission profile for a
sample of simulated clusters for a realistic choice of parameter
values (e.g., for the maximum shock acceleration efficiency, 
$\zeta_{p,max}$). In addition to the uncertainties in the model
parameters there is also uncertainty in the observationally derived
halo mass and gas density profile. In this work, we simply assume that
the shape of $q_{CR}(r,E)$ is given by the model described above and
account for the uncertainty in the model parameters as well as sample
variance with an additional normalization parameter, $\alpha_{CR}$, so
that
\begin{equation}
I_{CR}(r,E)=\alpha_{CR} \int_{l.o.s} \frac{q_{CR}(r,E)}{4\pi} dl.
\end{equation}
We take $\alpha_{CR}=1$ as our fiducial CR model and also consider the
case when $\alpha_{CR}$ is fitted from the actual gamma-ray data as an
optimal model. In the right panel of Figure~\ref{fig_profile} we
compare the CR profile for the fiducial model to the expected DM
annihilation profile within our three clusters, assuming a fiducial DM
particle model with particle mass $M\approx 100{\rm GeV}$ annihilating
through the \bbbar channel with cross-section, $\sigma
v=3\times10^{-26} \rm{cm^3 s^{-1}}$. In general the CR emission is
more centrally concentrated than the annihilation profile since the CR
traces the gas profile. It can be readily seen that Fornax has a
particularly low CR level while Coma is CR dominated. Coma has steeper
profiles due to its larger distance and hence smaller angular size.

\myfig
\includegraphics[width=0.5\textwidth]{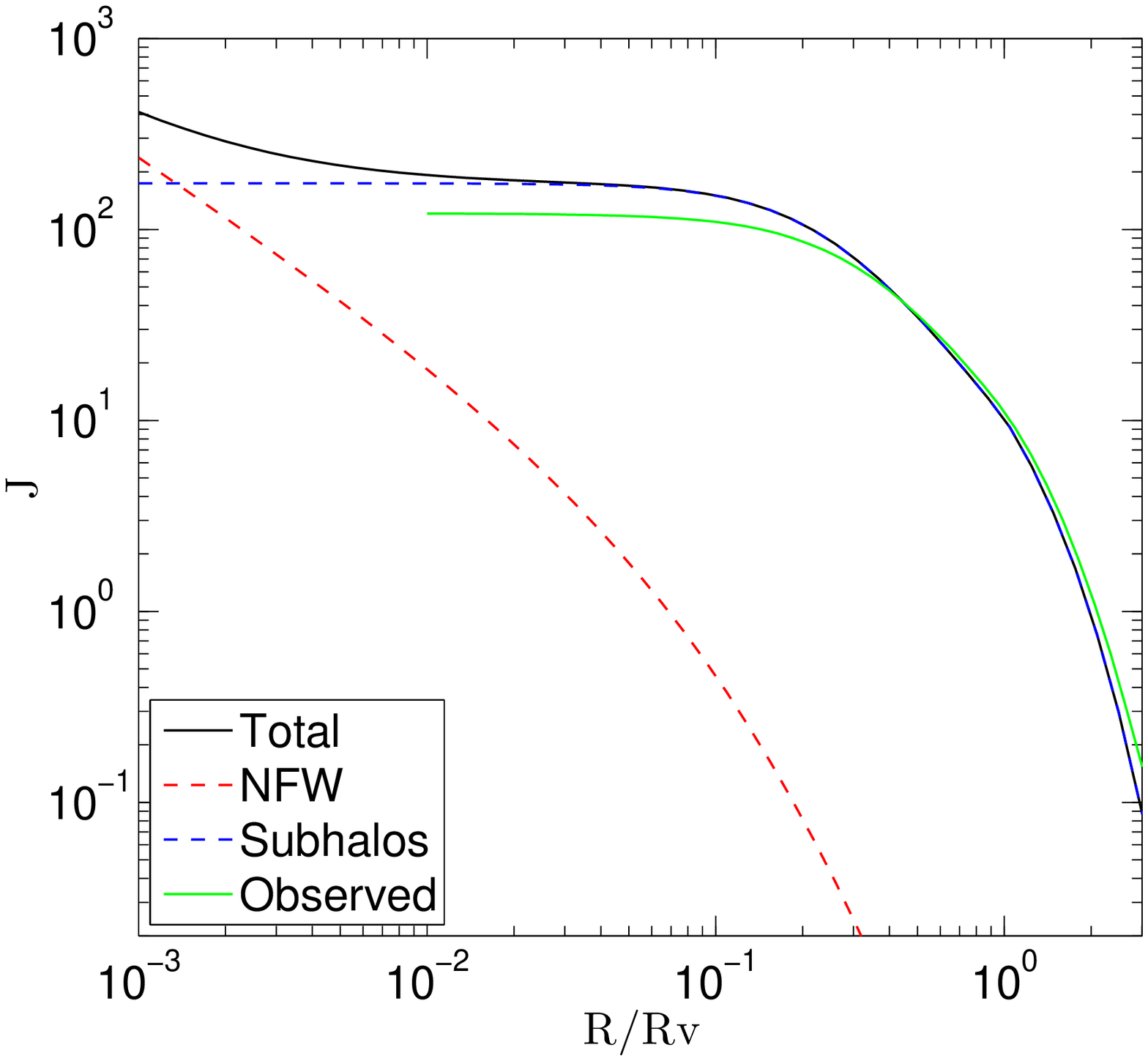}%
\includegraphics[width=0.5\textwidth]{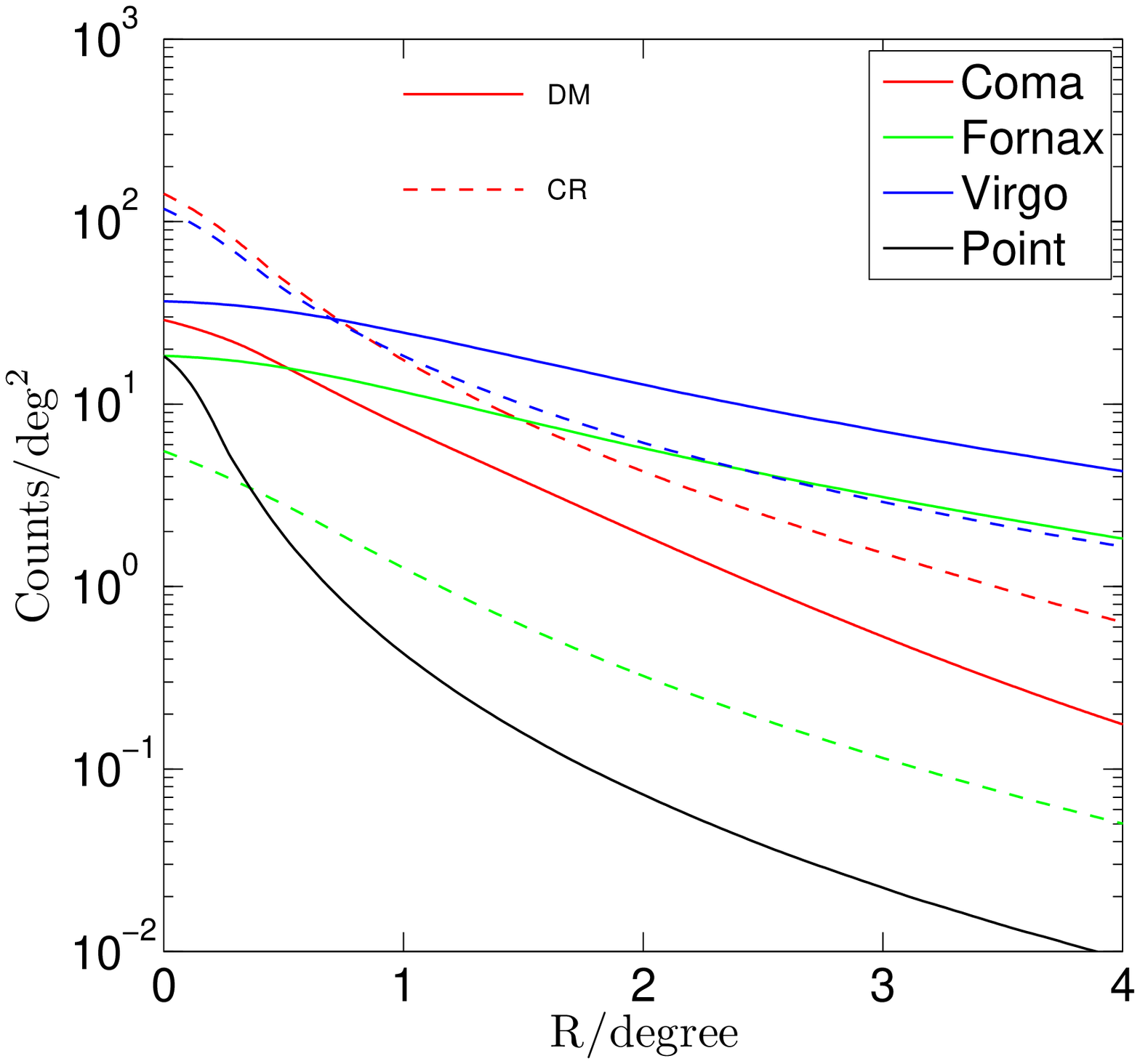}
\caption{Cluster photon profiles. Left: theoretical and PSF-convolved
J profile for Virgo. The total annihilation profile is shown as a
black solid line and is decomposed into the smooth main halo part (red
dashed line) and the subhalo part (blue dashed line). The green solid
line shows the annihilation profile after PSF convolution, plotted
down to an inner radius comparable to the pixel size of
0.1~deg. Right: PSF-convolved photon profiles from annihilation
(solid) and cosmic rays (dashed) for three clusters (indicated by
different colors). Solid lines show the expected photon count
profile for the extended DM annihilation model. Dashed lines show the
expected cosmic-ray induced photon counts for the fiducial CR
model. For comparison, we also plot the PSF-convolved profile for a
central point source model (black solid line) with arbitrary
normalization. In both panels, a dark matter model with particle mass,
$M\approx 100{\rm GeV}$, and annihilation cross-section, $\sigma
v=3\times10^{-26} \rm{cm^3 s^{-1}}$, through the \bbbar channel is
assumed.}\label{fig_profile}
\end{figure}

\section{Data analysis}
\subsection{Data preparation}\label{sec_prep}

We analyze the first 3 years of data (04/08/2008 to 16/08/2011) from
the
Fermi-LAT, \footnote{http://fermi.gsfc.nasa.gov/cgi-bin/ssc/LAT/LATDataQuery.cgi}
trimmed with the cuts listed below, to select high quality photon
events. This typically results in $\sim 10^5$ photons within a radius
of 10 degrees around each cluster, while the expected number of
annihilation photons is of the order of $10^2$ according to
Figure~\ref{fig_profile}. The most recent instrument response
function, P7SOURCE\_V6, is adopted for the analysis, in accordance
with our event class selection.

\begin{tabbing}
\hspace{5cm}\=\kill
Minimum Energy     \>    100 MeV\\
Maximum Energy     \>   100 GeV\\
Maximum zenith angle\footnotemark  \> 100 degrees\\
Event Class\footnotemark  \> 2 (P7SOURCE) \\
DATA-QUAL\footnotemark  \> 1 \\
LAT CONFIG\footnotemark  \> 1 \\
ABS (ROCK ANGLE)\footnotemark \> $<52$\\
ROI-based zenith angle cut \> yes \\
\end{tabbing}
 
\footnotetext[4]{ZENITH ANGLE (degrees): angle between the
reconstructed event direction and the zenith line (originates at the
center of the Earth and passes through the center of mass of the
spacecraft, pointing outward). The Earth's limb lies at a zenith angle
of 113 degrees.} 
\footnotetext[5]{EVENT CLASS: flag indicating the probability of the
event being a photon and the quality of the event reconstruction.}
\footnotetext[6]{DATA-QUAL: flag indicating the quality of the LAT
data, where 1 = OK, 2 = waiting review, 3 = good with bad parts, 0 =
bad} 
\footnotetext[7]{LAT-CONFIG: flag for the configuration of the lat (1
= nominal science configuration, 0 = not recommended for analysis)} 
\footnotetext[8]{ROCK ANGLE: angle of the spacecraft $z$-axis from the
zenith (positive values indicate a rock toward the north).} 

We list the basic properties of the three clusters in Table~\ref{table_property}. 
\begin{center}
\mytab
\caption{Basic Properties of Target Clusters}\label{table_property}
\begin{tabular}{|c|c|c|c|c|}
\hline   & Coma & Fornax & Virgo (M87) \\ 
\hline RA (deg) & 194.9468 & 54.6686 & 187.6958  \\ 
\hline DEC (deg) & 27.9388 & -35.3103 & 12.3369 \\ 
\hline $D_A$ (Mpc)\tablenotemark{a} & 95.8 & 17.5 & 22.3 \\ 
\hline $M_{200}$ ($\msun$)\tablenotemark{b} & 1.3e15 & 2.4e14 & 6.9e14 \\ 
\hline $r_{200}$ (deg)\tablenotemark{b} & 1.3 & 4.1 & 4.6 \\
\hline $\mathcal{J}_{NFW}$\tablenotemark{c} & 5.9e-5 & 4.1e-4 & 6.3e-4 \\
\hline Boost factor\tablenotemark{d} & 1.3e3 & 6.5e2 & 9.8e2\\
\hline
\end{tabular} \\
\tablenotetext{a}{Angular diameter distance from the NASA extragalactic database.}
\tablenotetext{b}{Cluster halo mass defined as mass within the radius,
$r_{200}$, within which the average density equals 200 times the
critical density of the universe. Values of masses taken from
\citet{Pinzke2}.} 
\tablenotetext{c}{Integrated coefficient,
$\mathcal{J}_{int}=\int_{\Delta \Omega} J d\Omega$, over the solid
angle spanned by the cluster virial radius, assuming a smooth NFW
density profile.} 
\tablenotetext{d}{Boost factor relative to the total annihilation
luminosity within the virial radius due to substructures, 
extrapolated to a subhalo mass limit of $10^{-6}\msun$} 
\end{table}
\end{center}
%\clearpage

\subsection{Maximum-likelihood fitting}\label{sec_ML}

We use the \texttt{pyLikelihood} tool shipped with the Fermi Science
Tools software package (version v9r23p1-fssc-20111006) to perform a maximum likelihood (ML) analysis
\citep{EGRET}. After applying appropriate data cuts, as described in
section~\ref{sec_prep}, we bin the data into 0.1 degree-wide pixels
and 30 logarithmic energy bins within a radius of 10 degrees around
each cluster. This large radius is chosen to account for the large LAT
PSF size ($4\sim10$ degrees at 100 MeV\footnote{The LAT PSF size scales
roughly as $E^{-0.8}$, so at 1~GeV it is $\sim 1$deg}).  

A model is constructed to fit the data including 
all known foreground and background emission, as well as  DM
and CR components. We include all the point sources from
2FGL within a radius of 15 degrees from the cluster center in the
model, plus the most recent galactic (GAL) and extragalactic (EG)
diffuse emission given by the template files
\texttt{gal\_2yearp7v6\_v0.fits} and \texttt{iso\_p7v6source.txt}. The
normalization of the GAL and EG diffuse components are allowed to
vary during the fitting. Within the cluster virial radius there are
two 2FGL point sources in Fornax, and three in Virgo, one of which is
associated with the central AGN, M87 \citep{M87Fermi}. We allow the
normalization and power-law spectral index of these five point sources
to vary freely. Parameters for other point sources are fixed as in the
2FGL catalog. 

A surface brightness template given by the dimensionless factor J
in Equation~\ref{eq_J} is generated for each cluster out to a 15
degree radius by summing up both the contribution from a smooth NFW
profile and the boost from subhalos. This J-map is used to fit for
extended cluster annihilation emission. For the point source 
model, the integrated factor $\mathcal{J}_{NFW}$ in Equation~\ref{eq_JNFW} is
used to derive an annihilation cross-section from the fitted total
flux.\footnote{This only strictly applies if  the cluster size is
smaller than the PSF size so that the entire cluster within the virial
radius can be approximated as a point source. Otherwise, the
integration over solid angle should only extend to the angular
resolution of the telescope. However, since the surface brightness
for a smooth NFW profile drops rapidly with radius, and because of the
energy dependent PSF size, we choose to use a generic estimation of
$J_{int}$ as given by an integral over the cluster virial radius. } 
Similarly, a CR photon template is generated for each cluster out to
three times the cluster virial radius, where the surface brightness
has dropped to below $10^{-5}$ of the central value and beyond which
the model is not reliable. Images for various model components are shown 
in Figure~\ref{f_components} taking Virgo as an example. 

The photon counts within each pixel are treated assuming Poisson
statistics for each energy bin to calculate the likelihood. The
best-fit parameters are obtained when the likelihood for the entire data set
is maximized. The significance of a given component of interest
(e.g. DM or CR) from the ML fitting is quantified by the likelihood
ratio statistic,

\begin{equation}
TS=-2\ln(L_0/L),
\end{equation}
where $L$ is the maximum likelihood for the full model, and $L_0$ is
the maximum likelihood for the null hypothesis, i.e, the model without
the component of interest. This test statistic (TS) approximately follows a $\chi^2$
distribution, with one degree of freedom for our case where the
normalization is the only free parameter. The probability that a given
value of TS arises purely from fluctuations of the null hypothesis is:
\begin{equation}
P=\int_{TS}^{\infty}\frac{1}{2}\chi_1^2(\xi)d\xi
=\int_{\sqrt{TS}}^{\infty}\frac{e^{-x^2/2}}{\sqrt{2\pi}}dx. 
\end{equation}
The factor $\displaystyle\frac{1}{2}$ comes from the constraint that
the normalization parameter is non-negative. The significance of
detection can thus be quoted as $\sqrt{TS}\sigma$ (one sided Gaussian
confidence).

\section{Results}
\subsection{Constraints on CR emission}

With all the model components defined above, we first proceed with ML
fitting for a model with no DM annihilation but with cosmic rays, the
``CR-only'' model hereafter. Note that the GAL and EG backgrounds, as
well as the nearby point sources are always implicitly included in the
analysis, as described in section~\ref{sec_ML}. The results for the
CR-only model fitting are listed in Table~\ref{table_CR}. The fitted
CR levels all agree within a factor of three with the theoretical
predictions. While Fornax is the most poorly constrained due to its
intrinsically low CR level, the derived upper limit for Coma already
rules out the fiducial value at 95\% confidence.
\begin{center}
\mytab
\caption{Fitting to the CR-only Model}\label{table_CR}
\begin{tabular}{|c|c|c|c|c|}
\hline  & $\alpha_{CR,fit}$\tablenotemark{a} & $\alpha_{CR,UL}$\tablenotemark{b} & $ F_{CR,UL}\tablenotemark{c}$~(ph $\cdot$ cm$^{-2}$ s$^{-1}$) & TS \\ 
\hline Coma & $0.3\pm 0.1$ & 0.5 & 2.74e-09 & 5.2 \\ 
\hline Fornax & $2.0\pm 2.6$ & 6.4 & 2.4e-09 & 0.6 \\ 
\hline Virgo & $1.5\pm 0.5$ & 2.3 & 2.1e-08 & 8.2 \\ 
\hline 
\end{tabular}\\ 
\tablenotetext{a}{Best fit normalization ($\alpha_{CR,fit}=1$ is 
the theoretical prediction.)} 
\tablenotetext{b}{95\% upper limit (UL) on the normalization}
\tablenotetext{c}{95\% upper limit on the CR induced gamma-ray flux
from 100~MeV to 100~{\rm GeV}} 
\end{table}
\end{center}
%\clearpage
\subsection{Constraints on DM annihilation}

Given the low significance of the CR detection in the CR-only model, it is not
safe simply to adopt the best fit $\alpha_{CR}$ values for further
extraction of the DM signal. Instead, we consider the following
four families of cosmic ray models in the presence of a DM component: 

\begin{description}
\item[Fiducial-CR model.] The CR level is fixed to the theoretical
expectation, $\alpha_{CR}=1$. Since this  is larger than the  upper
limits derived for Coma, we exclude Coma from further discussion for this family. 
\item[Optimal-CR model.] The  CR level is taken as  the best-fit value listed in
Table~\ref{table_CR}.  
\item[Free-CR model.] The normalization of the CR level is treated as
a free parameter.
\item[No-CR model.] No CR emission is considered, only DM.
\end{description}
For each family, both point source (PT) and extended (EXT) profiles
are considered for the DM component.

\subsection{The \bbbar channel}

The significance for the \bbbar channel of annihilating DM is shown in
Figure~\ref{fig_TSbb}. It can be seen that in almost all cases, the
EXT profile has higher significance than the PT profile, over the
entire DM mass range considered. There is no significant evidence for
a PT model component within Fornax. For the four families of models,
the No-CR model has the highest DM significance. In particular, a peak
TS value of 19.8, corresponding to a $4.4\sigma$ confidence level, is
obtained for Virgo with the EXT profile when no CR component is
considered. This decreases to $3.4\sigma$ when a CR component with
free normalization is included and is even lower when the CR level is
fixed either to the fiducial or the optimal levels. There is a general
trend for the significance to peak in the mass range $20-60$~GeV for
all three clusters. This is consistent with the conclusion of
\citet{GalCen}, who claim that a DM model with particle mass in the
range $25-45$~GeV annihilating into \bbbar final states can explain
the excess extended emission observed from the direction of the
Galactic center.
\myfig
\includegraphics[scale=0.7]{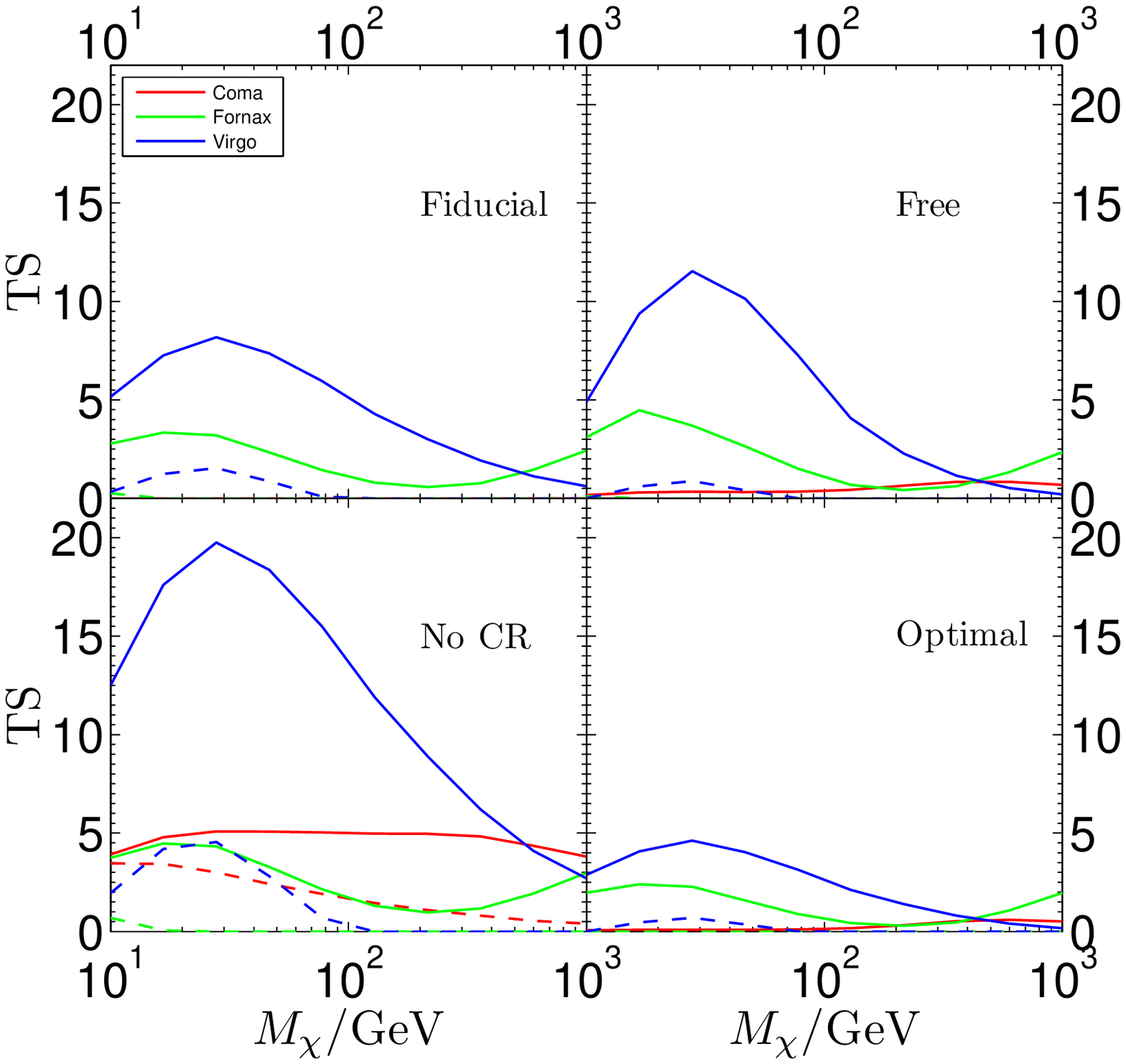}
\caption{TS values for a DM component annihilating through the \bbbar
channel. Different colors represent different clusters. Solid lines
correspond to the extended model while dashed lines correspond to the
point source model. The four panels are for the various CR models as
labeled.}\label{fig_TSbb} 
\end{figure}
%\myfig
%\includegraphics[width=0.3\textwidth]{LikelihoodComa.BBbarIC.eps}%
%\includegraphics[width=0.3\textwidth]{LikelihoodFornax.BBbarIC.eps}%
%\includegraphics[width=0.3\textwidth]{LikelihoodVirgo.BBbarIC.eps}
%\caption{Likelihood for various model fittings. Plotted are the
%logarithm of the maximum likelihood values offset by that for the
%No-CR model without DM component, as a function of adopted DM particle
%mass, for Coma, Fornax and Virgo from left panel to right panel. Four
%colors correspond to the four families of models regarding CR
%freedom. For each model family, the solid line shows the likelihood
%when an extended annihilation component is adopted, the dashed line
%shows that when point source annihilation component is adopted, and
%the dotted line with circles shows the likelihood for the model with
%no annihilation emission.}
\label{fig_Likebb} 
%\end{figure}
\myfig
\includegraphics[width=0.5\textwidth]{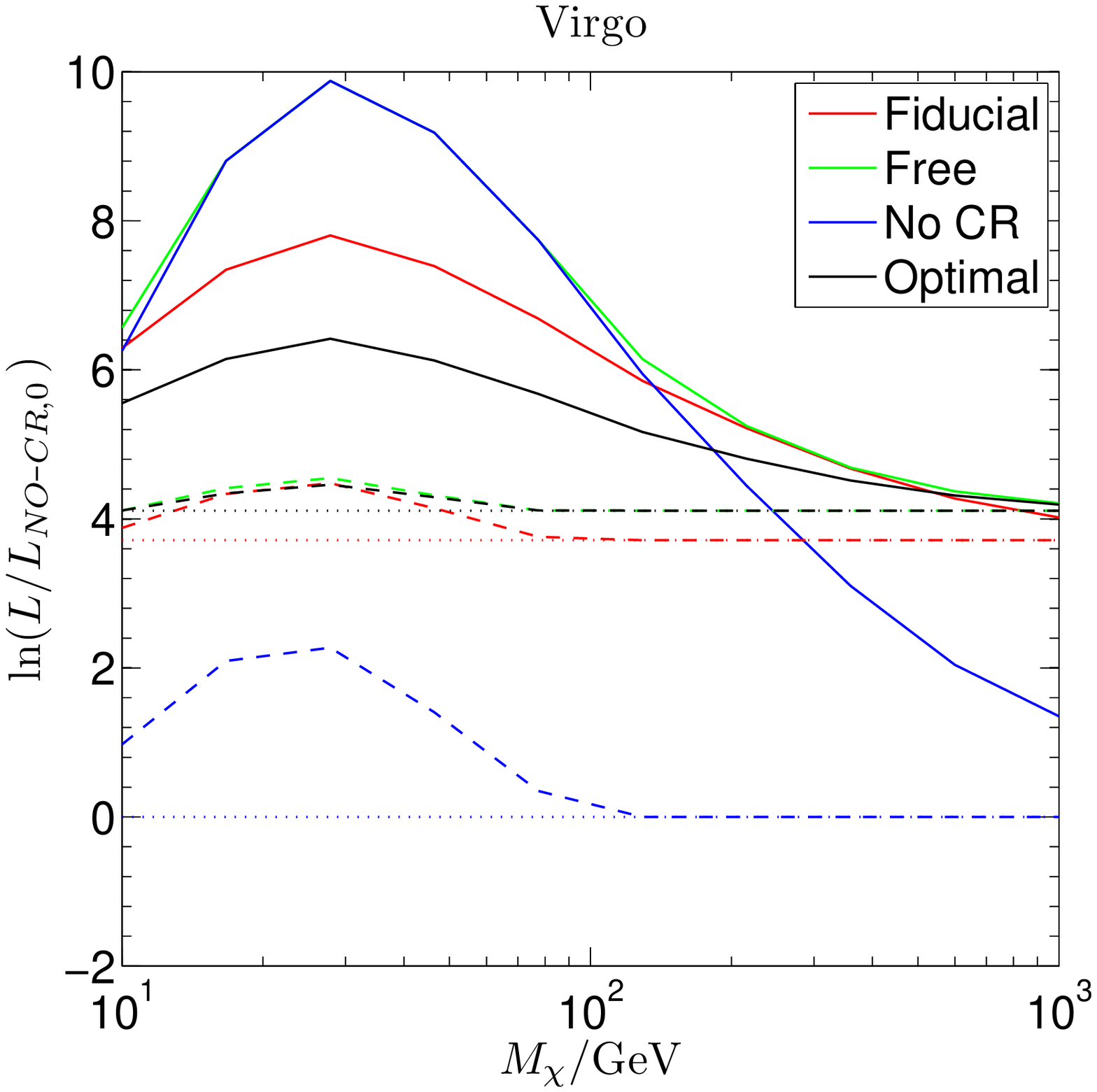}
\caption{Likelihoods for various model fits to the Virgo cluster
data. Plotted is the logarithm of the maximum likelihood values
relative to that for the no-CR model with no DM component, as a
function of the DM particle mass. The colors correspond to the four
families of CR models. For each model family, the solid line shows the
likelihood when the EXT profile is assumed, while the dashed line
shows the likelihood for the PT approximation. The dotted lines show
the likelihood for models with no annihilation emission, in which case
the Free-CR and Optimal-CR models are equivalent.}
\label{fig_Likebb}
\end{figure}

To facilitate comparison between all different model families and
assess the significance of a CR component, we show the maximum
likelihood values for each model of the Virgo cluster in
Figure~\ref{fig_Likebb}. As seen in the TS curves, the likelihood for
the extended model is always higher than that for the point source
model. For the most likely mass range, $20\sim60$~GeV, the No-CR and
Free-CR models with extended DM emission share the highest likelihood,
with the No-CR model being superior by having one fewer parameter.
Actually there is effectively no contribution from
CR when CR and DM are fitted simultaneously for this particle mass
range, and the fiducial and optimal CR levels are above the 95\% CR
upper limit predicted from the Free-CR model in the presence of
extended DM annihilation. This preference for DM over CR also obtains
for Fornax, although the optimal and fiducial CR cases are still
allowed in the presence of an extended DM component.  The story is
very different in Coma, where the DM and CR components are almost
degenerate in the sense that either component alone has a similar
level of likelihood.
 
In Figure~\ref{fig_Fluxbb} we show the 95\% confidence upper limits on
the DM annihilation flux and compare them to the CR levels. For each
cluster, the colored stripes are defined by the minimum and maximum
upper limit predictions among the four families of models. The optimal
CR levels in the three clusters are all comparable to the fitted DM
flux, and the DM flux upper limits for the four different CR models
vary only within a factor of two, with the No-CR and Optimal-CR cases
predicting the highest and lowest upper limits. The left and right
panels show the results for the PT and EXT models respectively; the PT
models always have lower flux upper limits than the extended models.

The flux upper limits are translated into cross-section upper limits
in Figure~\ref{fig_SigVbb}, using Equation~\ref{eq_I}. These are also
shown as colored regions reflecting the variance in the different
treatments of CR. Although the predicted flux upper limits decrease
slowly with DM particle mass and remain within the same order of
magnitude for the mass range considered, the resulting cross-section
upper limits increase by a factor of 100 from low to high particle
mass.  This is because low mass particles correspond to higher DM
number densities (the $\rho_\chi^2/M_\chi^2$ factor in
Equation~\ref{eq_I}) for a given mass density, so to obtain the same
flux level, the required cross-section must be smaller for low mass
particles. With the help of a boost factor of order $10^3$, a much
lower cross-section is needed (by a factor of at least 100) for
extended annihilation models to achieve a slightly higher flux upper
limit than point source models. 

Our cross-section limits are much lower than the 11-month Fermi-LAT
constraints in \citet{FermiCluster}, where the tightest constraint
came from Fornax for a much lower assumed boost factor of $\sim
10$. Our limits are also tighter than that from a joint analysis of
the dwarf satellites of the Milky Way by
\citet{Dwarf}. They drop below the fiducial thermal cross-section of
$3\times 10^{-26}{\rm cm}^3\rm{s}^{-1}$ for $M_\chi<100$GeV. Of the three
clusters, Virgo has the highest significance and flux upper limits and
places the tightest constraints on the annihilation cross-section. The best-fit cross-section
for the extended model is closer to the thermal cross-section than that for the point source model.
%Because the significance for the fitted DM component is in general still low, we do not give the best fit flux or cross-section for the entire mass range. Instead, we just quote the best-fit cross-section for $M_\chi=28$~GeV and when a TS value above 2 is found, which is $(3-6)\times 10^{-27}\rm{cm^3 s^{-1}}$ for the extended model and $(2-4)\times 10^{-24} \rm{cm^3 s^{-1}}$ for the point source model, for the three clusters. 
For example, for $M_\chi=28$~GeV and where the TS value is higher than 2,
the best-fit cross-section is $(3-6)\times 10^{-27}\rm{cm^3 s^{-1}}$ for the 
extended model and $(2-4)\times 10^{-24} \rm{cm^3 s^{-1}}$ for the point source model.

\myfig
\includegraphics[width=0.5\textwidth]{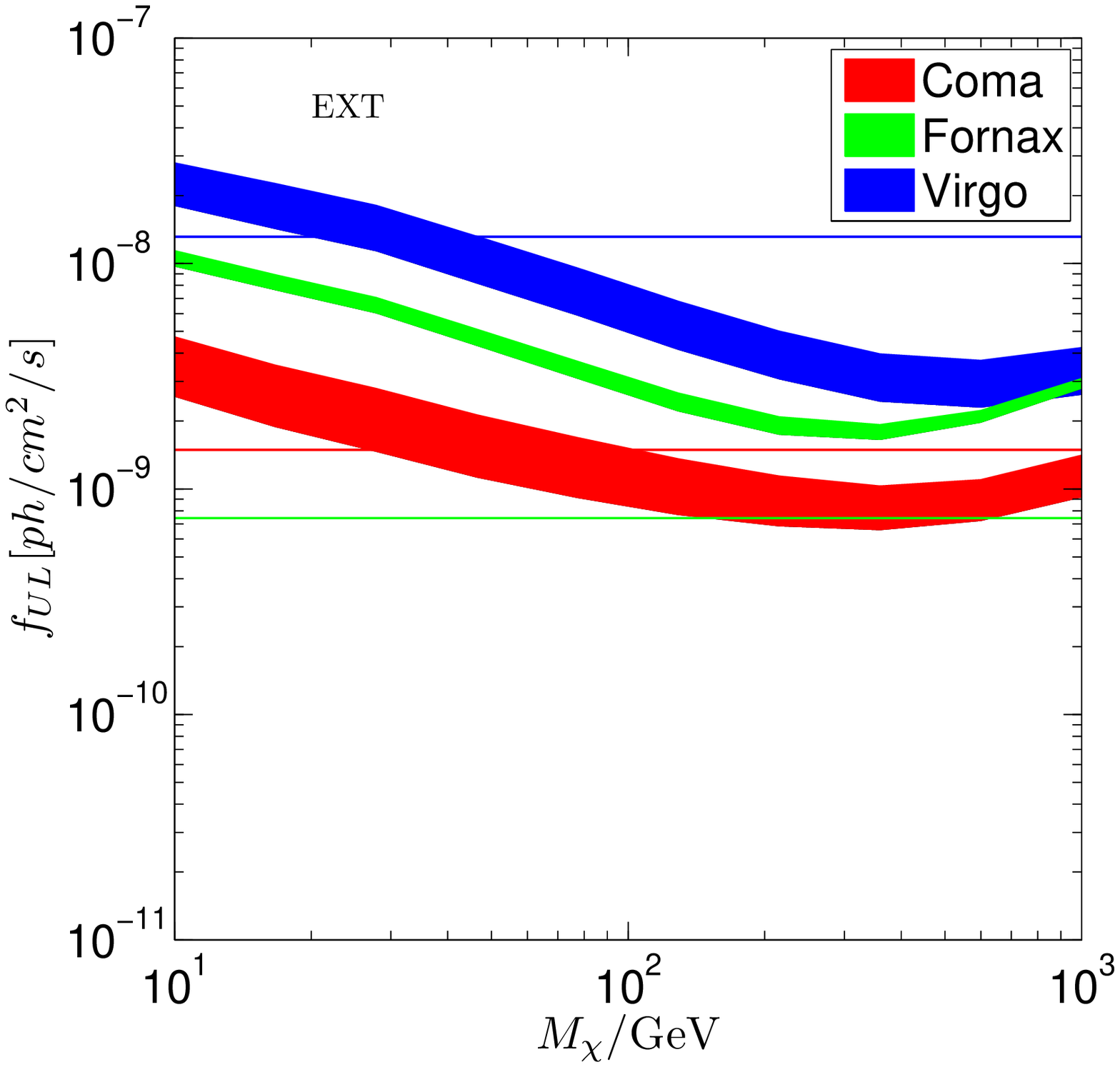}%
\includegraphics[width=0.5\textwidth]{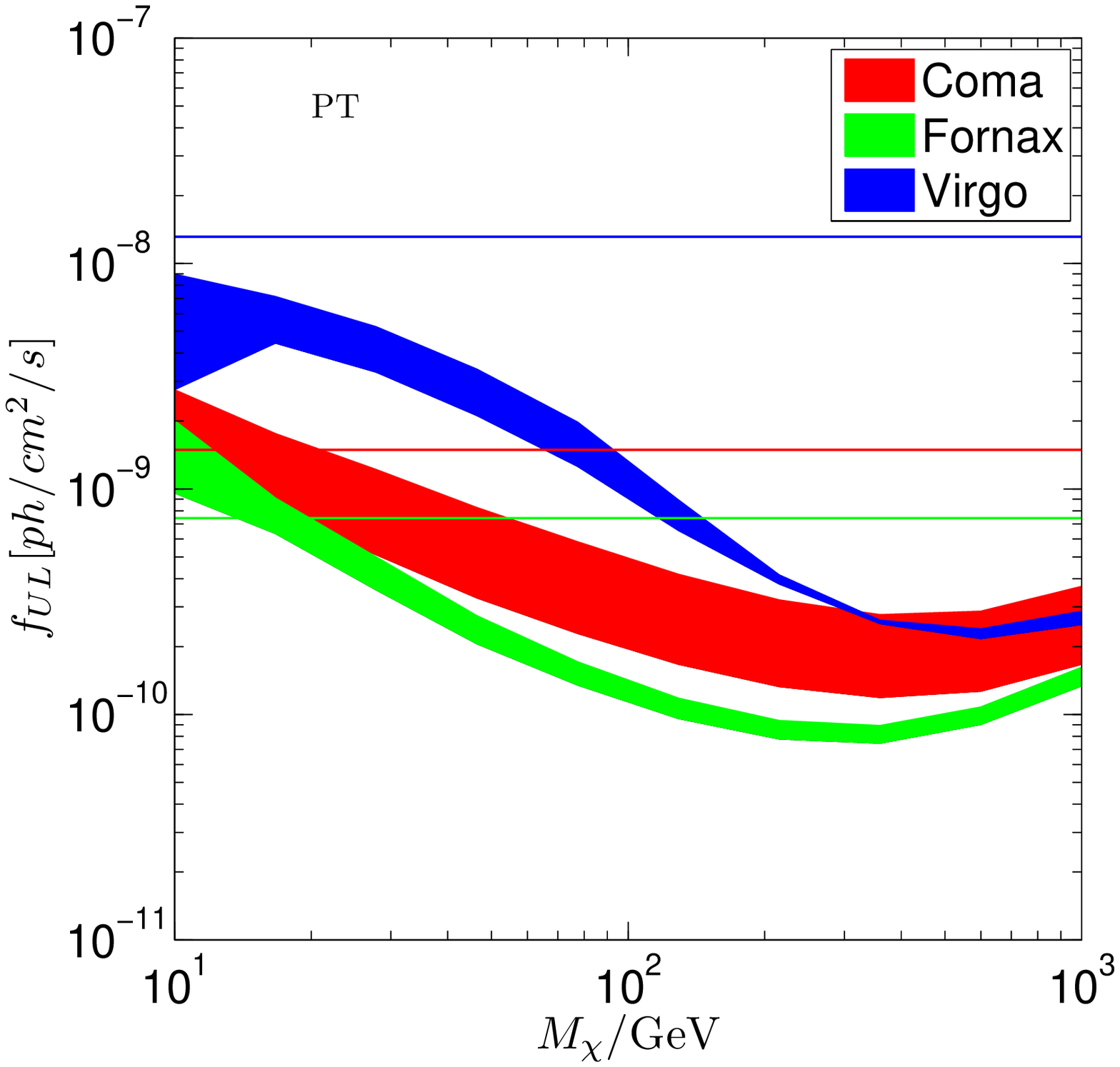}
\caption{DM annihilation flux upper limit for the \bbbar channel. The
stripes are defined by the minimum and maximum upper limits given by
the four CR model families, with different colors corresponding to
different clusters, as indicated in the legend. Left and right panels
are the results for the EXT and PT profiles respectively. For each
cluster, a solid line of the corresponding color shows the optimal CR
flux. }\label{fig_Fluxbb}
\end{figure}
\myfig
\begin{minipage}{\textwidth}
\includegraphics[scale=0.7]{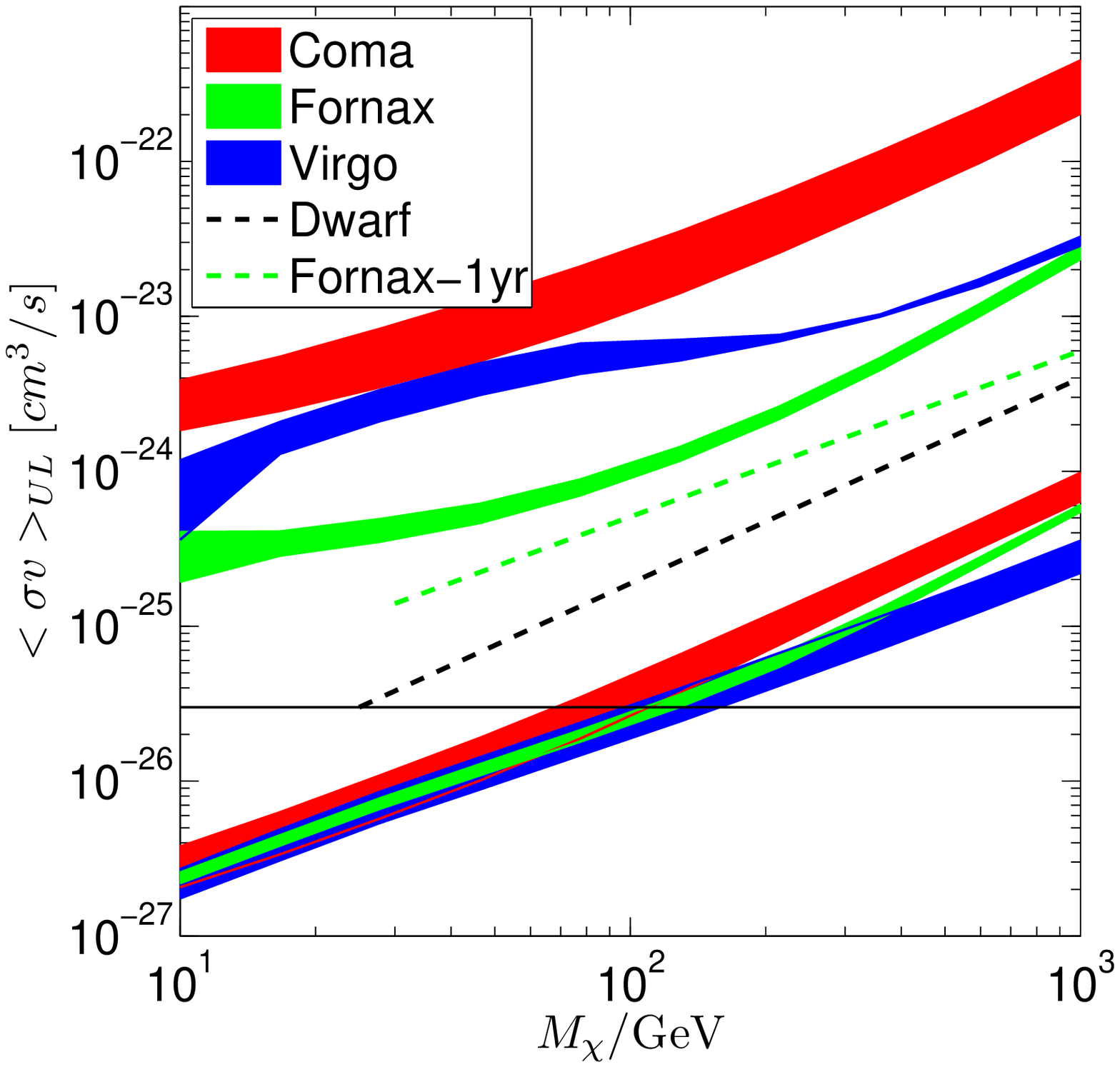}
\caption[]{Upper limit for the DM annihilation cross-section in the
\bbbar channel. The different colors represent the three clusters, with
the stripes spanning the range between the minimum and maximum upper
limits given by the four different ways of treating the CR
component. The three highest stripes show the PT model constraints and
the three lowest the EXT model constraints. We also plot with dashed
lines constraints\footnote{The lines for ``Dwarf" and ``Fermi-1yr" are
only reproduced schematically, by reading out several data points from
the original plots in the references.} from a joint analysis of the
Milky Way dwarf galaxies \citep[][black dashed line]{Dwarf} and
previous constraints from the 11-month Fermi-LAT data for Fornax
\citep[][green dashed line]{FermiCluster} assuming these authors' optimistic
value for the boost factor from subhalos, which gives the tightest
constraint. The black solid line indicates the canonical thermal
cross-section of $3\times 10^{-26} \rm{cm^3 s^{-1}}$. }\label{fig_SigVbb}
\end{minipage}
\end{figure}

\subsection{The \mumu channel}

The significance, flux upper limits and cross-section upper limits
for DM annihilating through the \mumu channel are plotted in
Figures~\ref{fig_TSmm} to~\ref{fig_SigVmm}. The predicted flux upper
limits for Coma and Virgo are still comparable to the CR level, with
Fornax having much lower CR emission. A DM particle mass of $2-10$~GeV
is preferred, which is also consistent with the values inferred from
analysis of the Galactic center emission by \cite{GalCen}. In
addition, a second region of high significance is obtained for
$M_\chi>1$~TeV. The inferred cross-section falls below the canonical
value for DM particle masses less than 10~GeV. Note the discontinuity
in the upper limit predictions around 100~GeV which reflects the
transition from the prompt annihilation dominated regime to the IC
emission dominated regime in the photon spectrum.

\myfig
\includegraphics[scale=0.7]{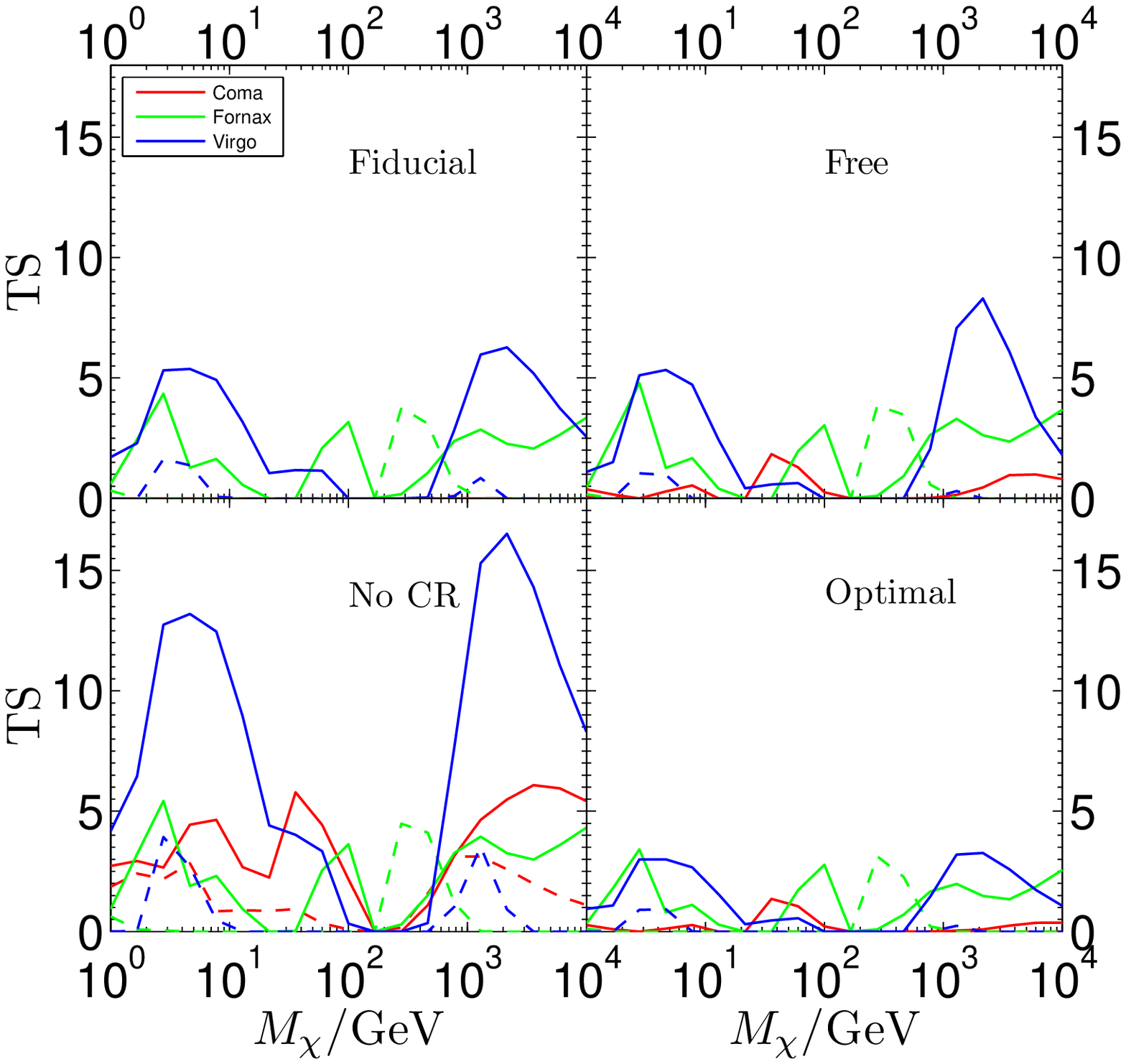}
\caption{TS values for DM annihilating through the \mumu
channel. Line styles are as in Figure~\ref{fig_TSbb}.}
\label{fig_TSmm} 
\end{figure}

\myfig
\includegraphics[width=0.5\textwidth]{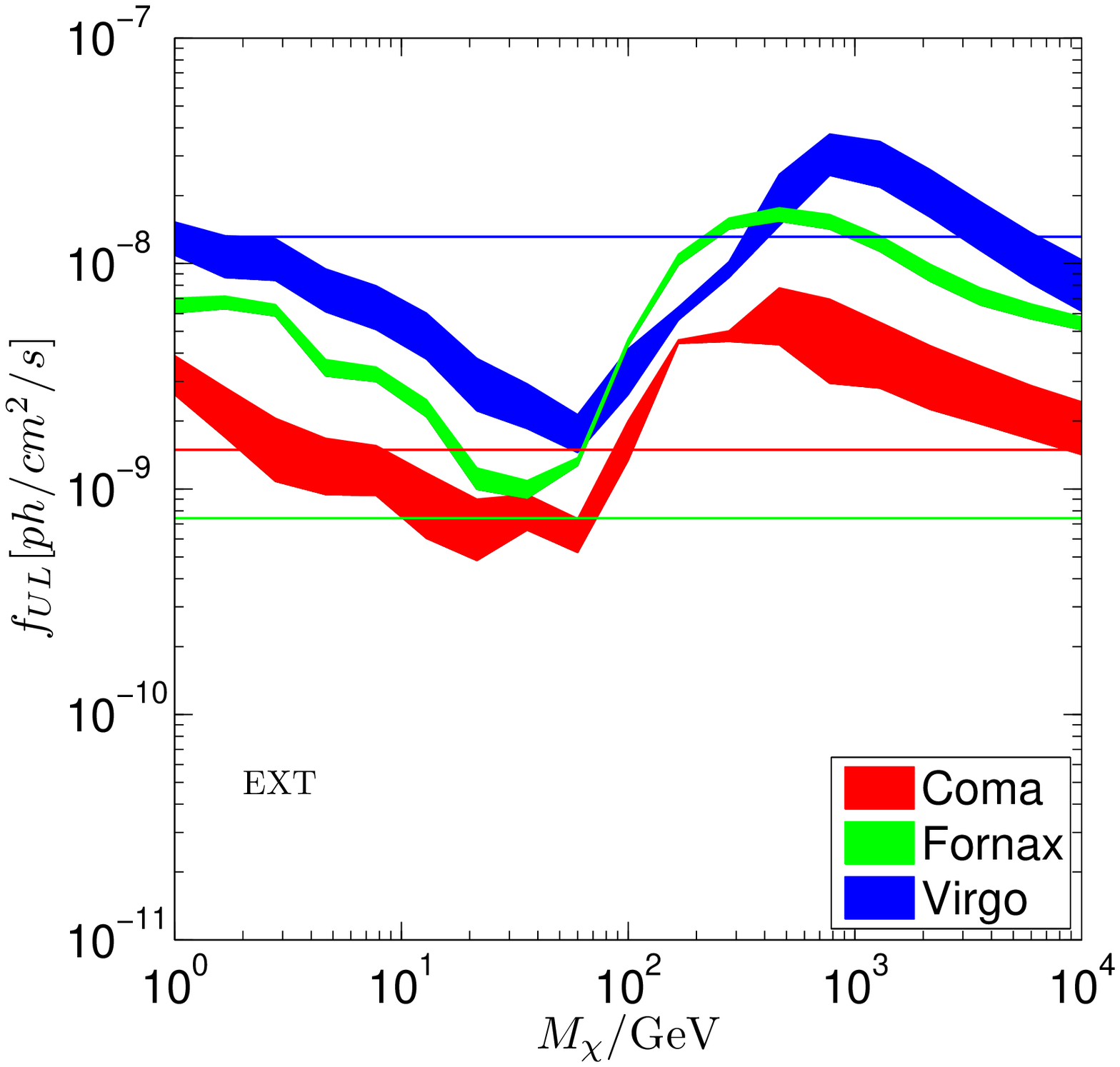}%
\includegraphics[width=0.5\textwidth]{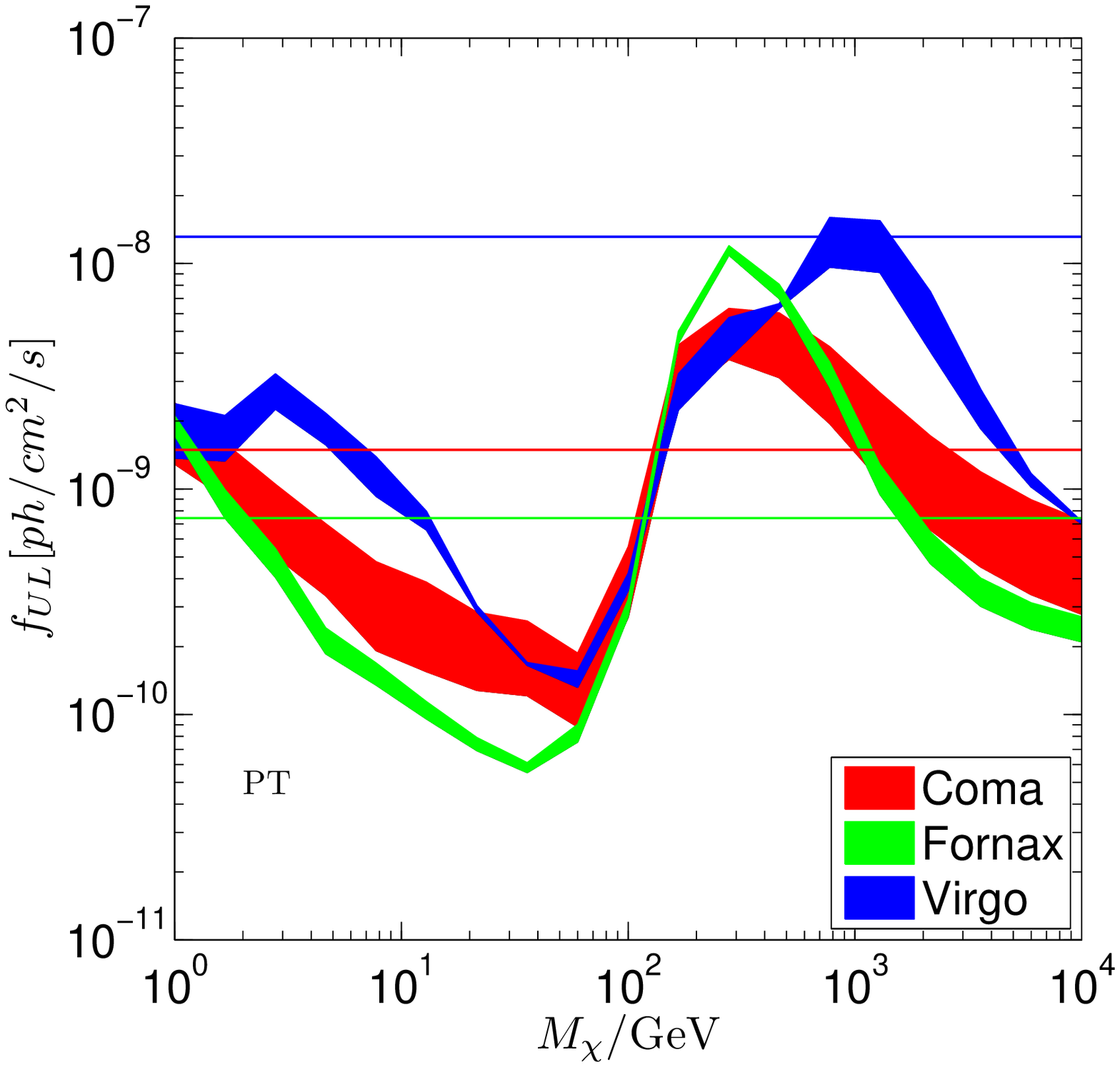}
\caption{DM annihilation flux upper limits in the \mumu channel. Line
styles are as in Figure~\ref{fig_Fluxbb}.}
\label{fig_Fluxmm}
\end{figure}

\myfig
\includegraphics[scale=0.7]{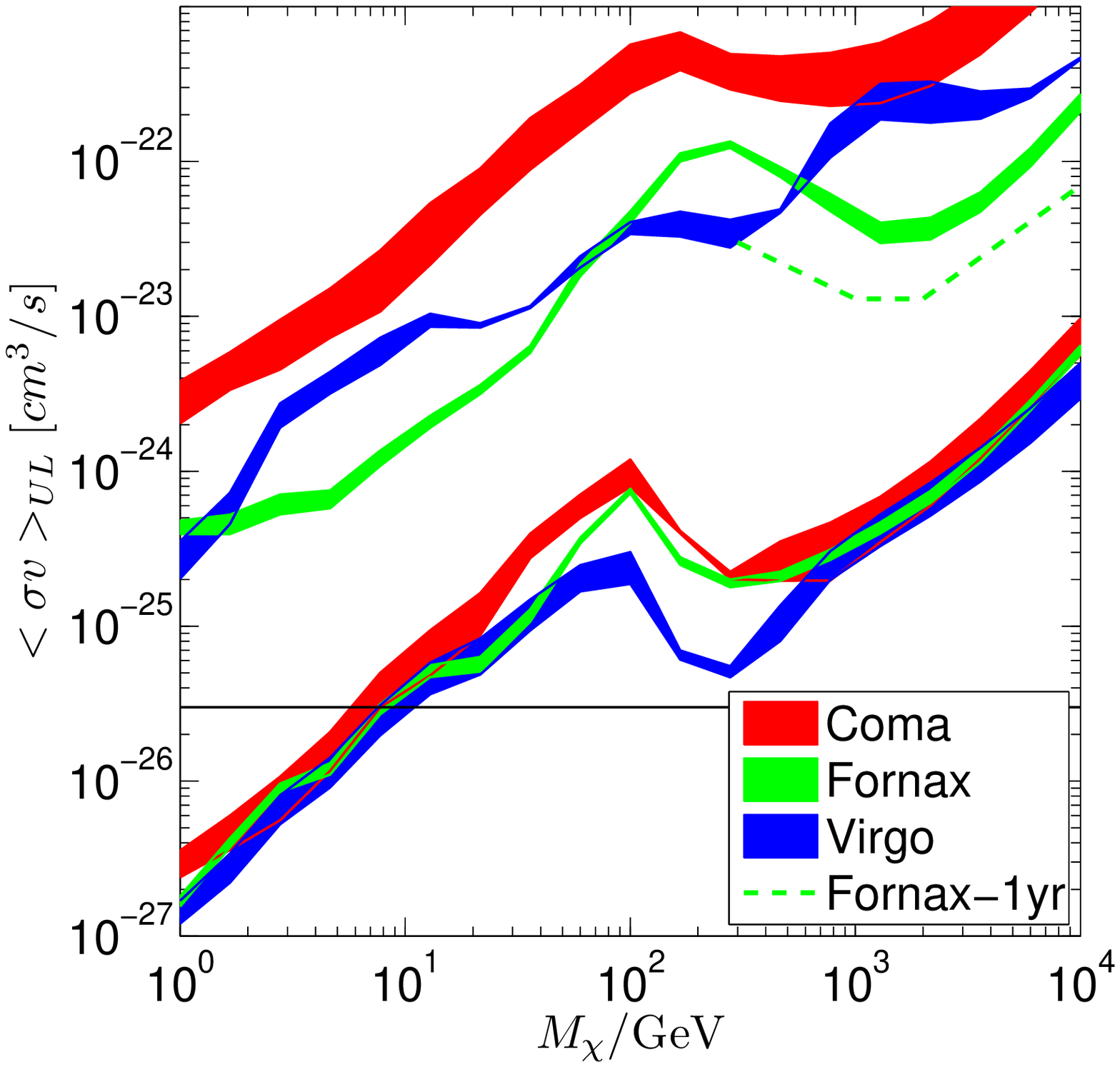}
\caption{Upper limits for the DM annihilation cross-section in the
\mumu channel. Line styles are as in Figure~\ref{fig_SigVbb}.
The green dashed line is the 11-month
Fermi result \citep{FermiCluster} for Fornax.}
\label{fig_SigVmm}
\end{figure}

\subsection{Examination of the excess emission}

To pin down the source of the significance of our fits, in
Figure~\ref{f_TSprof} we break down the TS value for our preferred
particle model into contributions from four radial ranges. Clearly,
most of the significance for the DM model comes from the region within
3~degrees around each cluster. The spectrum within this region
(Figure~\ref{f_SpecFit}) shows excess emission around 1~GeV for all
three clusters relative to a fit with no DM or CR components (the null
model). This excess is consistent with an almost scale-free spectrum
($dN/d{\rm ln}E\sim {\rm const}$), picking out DM particles in the range
$20-60$~GeV for \bbbar and $2-10$~GeV for \mumu final states, for
which the cutoff at the high energy end and the IC boost at the low
energy end are not significant around GeV energy scales. At much
higher masses (above 1~TeV) in the
\mumu channel, where IC emission completely dominates the annihilation
spectrum and begins to harden again around GeV scales, the TS for the 
annihilation component picks up again, as seen in
Figure~\ref{fig_TSmm}. 

The radial profile of the photon counts also reveals signs of excess
emission in the central regions of the clusters when fitted with the
null model, as shown in Figure~\ref{f_RadialFit}. The central cluster
emission is apparent in Coma and Fornax but in Virgo it can be
observed only weakly in a ring of [1,3] degrees, due to the existence
of a central point source, M87. We show the fitted photon counts
spectrum of M87 in the null, DM-only and CR-only models explicitly in
Figure~\ref{fig_AGNspec}. The total flux for M87 is comparable to the
CR and DM flux and varies by 20 to 30 percent in different models
without much change in the spectral shape. The low energy dip for the
three components comes from the decrease in the instrument's effective
area at low energy. The AGN spectrum differs from the CR spectrum
primarily at low energy and is distinct from the DM spectrum over a
wide energy range. This, together with their different radial
profiles, helps to disentangle the two components easily during the
fitting. In Fornax, however, the central excess is only partly fitted
by a DM component. We checked that the remaining part is mostly
associated with a low energy spectral excess below 400MeV, which is
not fitted by the current DM spectrum.

A possible reason for a central excess we have found in all three
clusters is a systematic error in the fitting. For instance, if the
fixed point sources outside the cluster virial radius have too much
flux attributed to them, then the diffuse EG and GAL components would
be biased low and would not contribute enough model flux to the
central region. This would lead to an excess in the center. However,
we have verified that our fits are insensitive to whether all the
point sources are kept fixed or allowed to vary freely.\footnote{When
the M87 parameters are fixed to the 2FGL values, the significance of
the DM detection in Virgo decreases to less than $3\sigma$. However,
we find that the 2FGL parameter values tend to overfit the current
data slightly. When the M87 parameters are refitted within the null
model, the normalization decreases by 15 percent and if this value is
then fixed when fitting for an additional DM component, the
significance for the DM and CR components are almost the same as when
the fits are performed in our standard way.} In addition, we have
carried out the same analysis on three regions randomly selected from
the 3-year LAT sky where no known prominent clusters exist near the
center. No central excess is observed in any of these three regions,
suggesting that the excess emission is related to the presence of the
galaxy clusters.

To see how common the extended emission is in galaxy clusters, we have
searched in seven other clusters adopting our preferred DM particle
model with $M_\chi\approx28$~GeV in the \bbbar channel. These seven
clusters -- M49, A1367, AWM7, Centaurus, Hydra, NGC4636 and NGC5813 --
are predicted to have high annihilation flux by
\citet{Pinzke2}. They are also selected to lie at least 15 degrees
away from the galactic plane. A DM significance comparable to that in
Fornax is found in M49 and A1367\footnote{The central excess in A1367
is more consistent with a point source profile, with a significance of
$3.5\sigma$.} with a TS value around 4, while Centaurus, NGC4636 and
NGC5813 show no sign of DM annihilation emission at all. We point out
that these additional clusters are all very compact on the sky (except
M49) with a virial radius close to or smaller than 2 degrees, making
it difficult to exploit the information contained in the radial
profile. In addition, the existence of additional sources inside the
virial radius (one point source in Hydra and NGC4636, two point
sources and one extended source (CenA Lobe) inside Centaurus)
complicates the analysis further. 

\myfig
\includegraphics[width=0.6\textwidth]{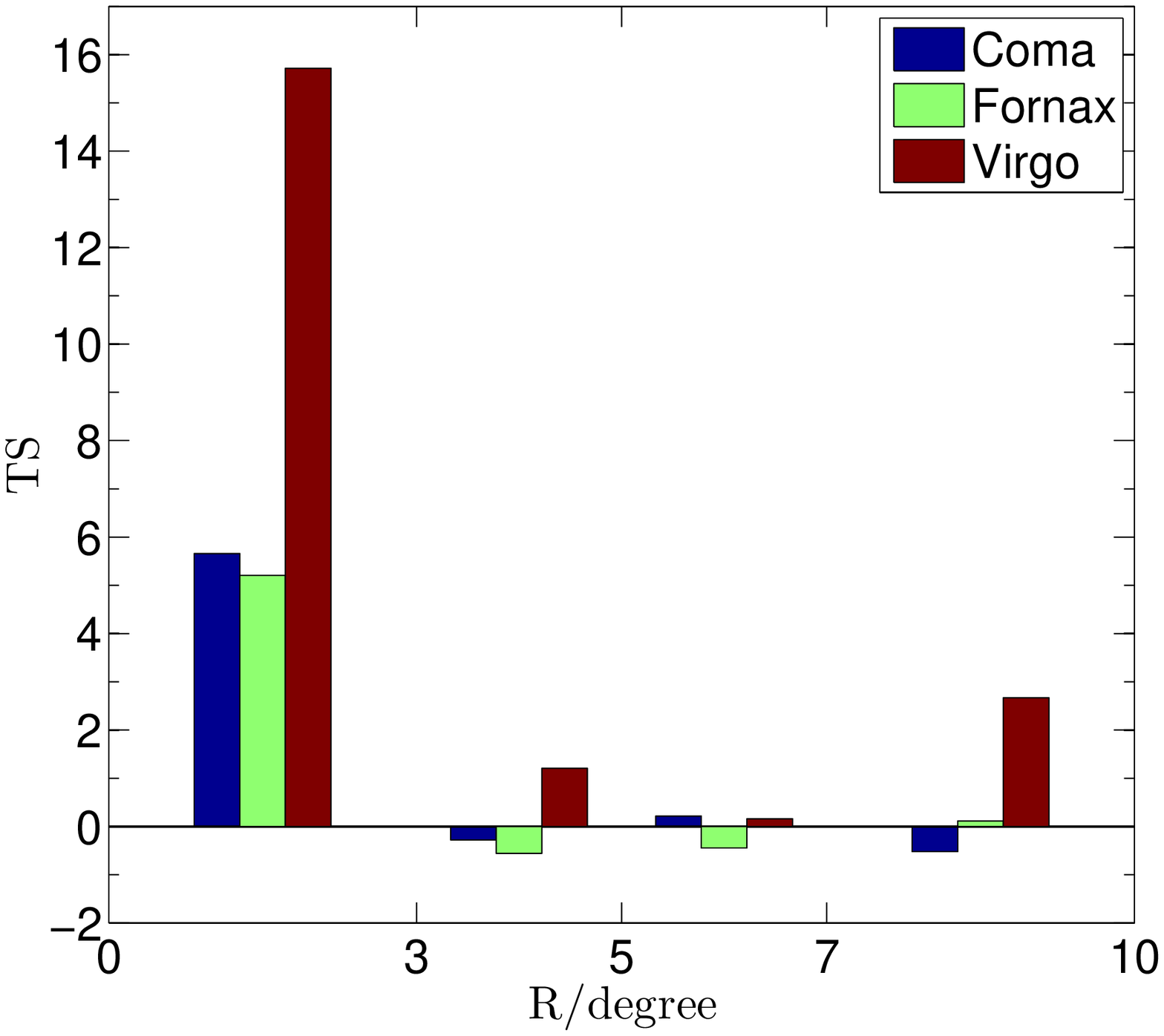} 
\caption{TS profile for our three clusters in the case of a DM model with
$M_\chi\approx 30$~GeV, annihilation through the \bbbar channel and an
extended emission profile. The TS value for each cluster is broken
down into contributions from each of four radial ranges,
[0,3],[3,5],[5,7] and [7,10] degrees, by calculating the likelihood
ratio inside each region for the global best fit parameters. The 
colors correspond to the three clusters as labelled. The significance
is dominated by regions within 3 degrees around each cluster
center.}\label{f_TSprof}
\end{figure}

\myfig
\includegraphics[width=0.33\textwidth]{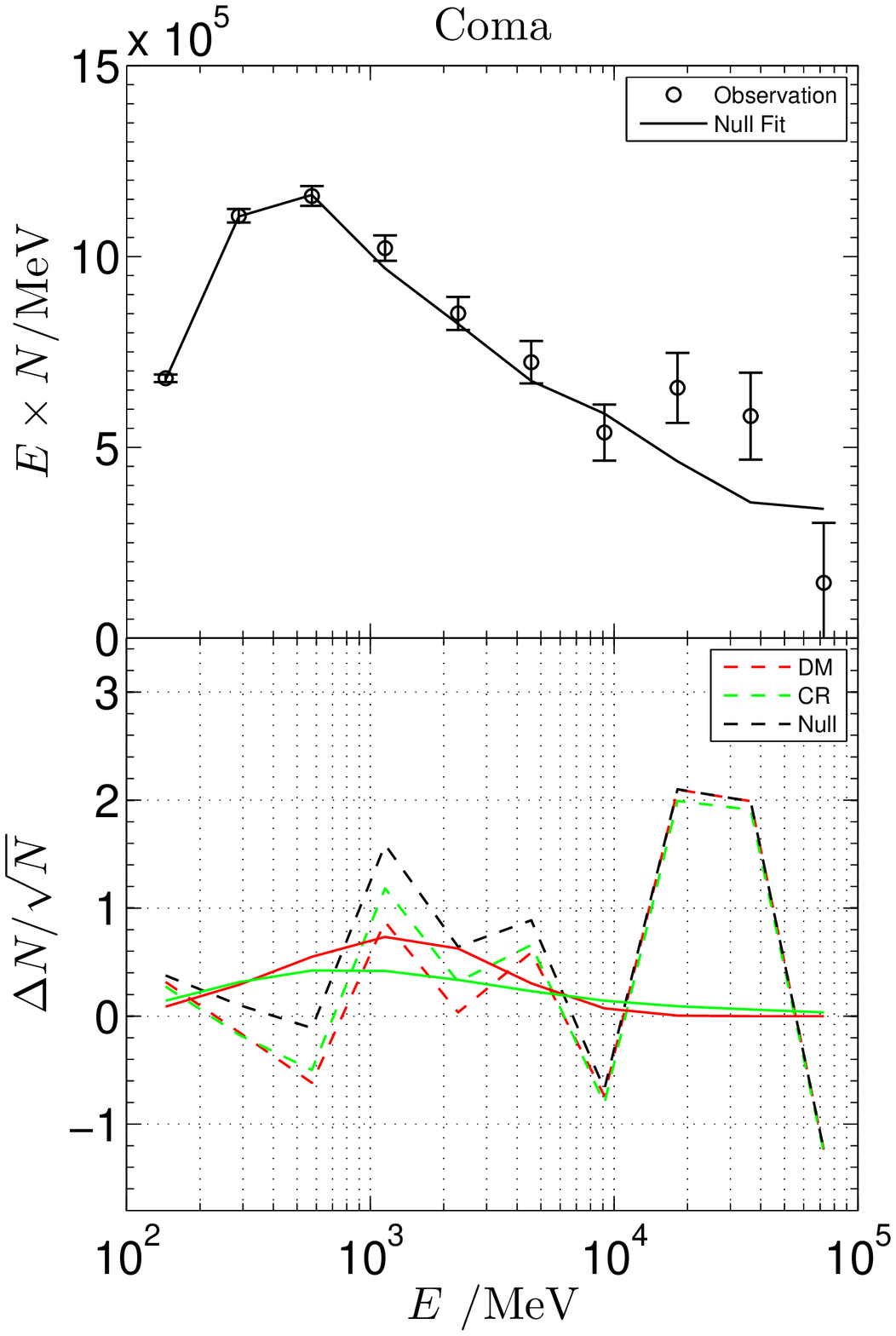}% 
\includegraphics[width=0.33\textwidth]{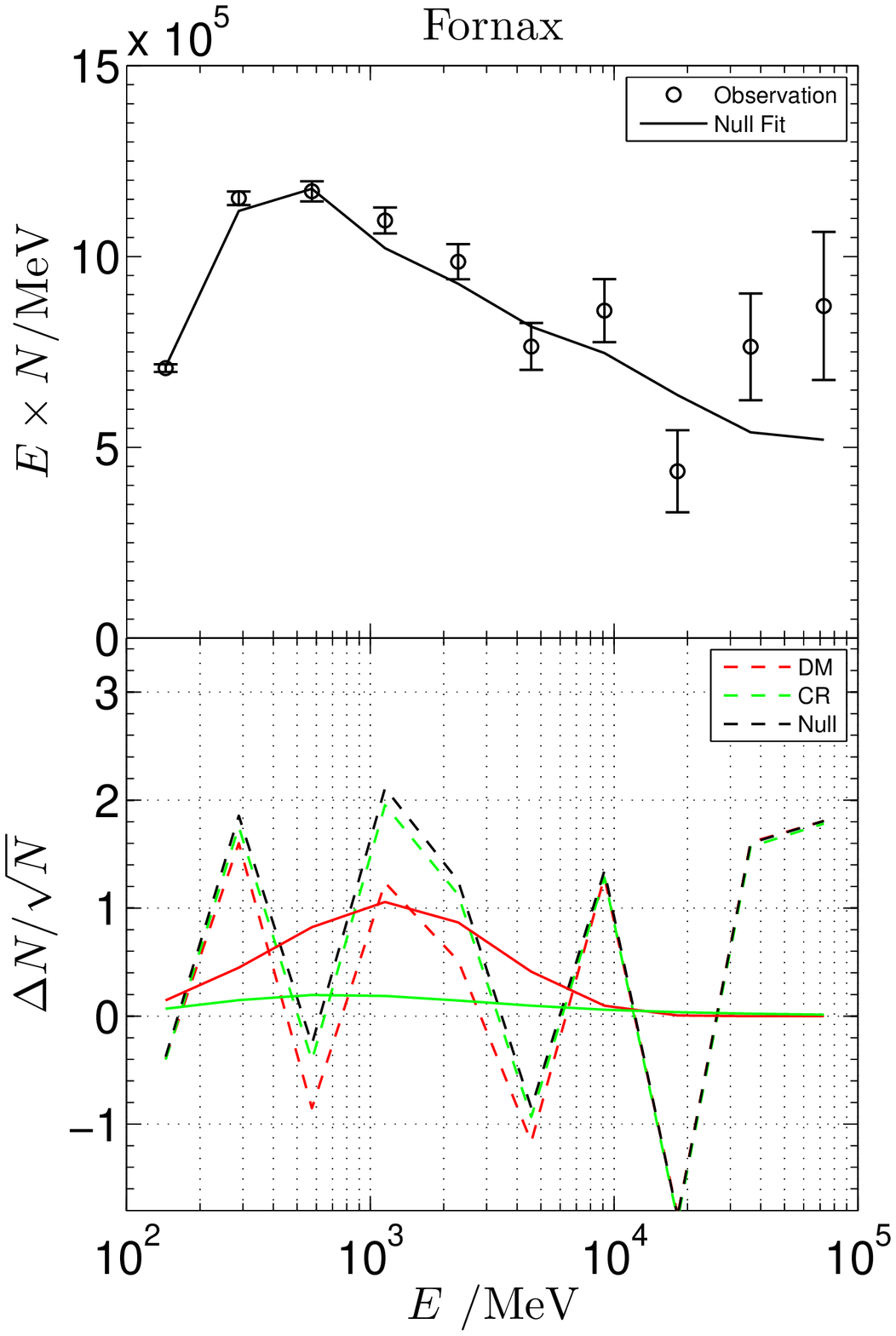}% 
\includegraphics[width=0.33\textwidth]{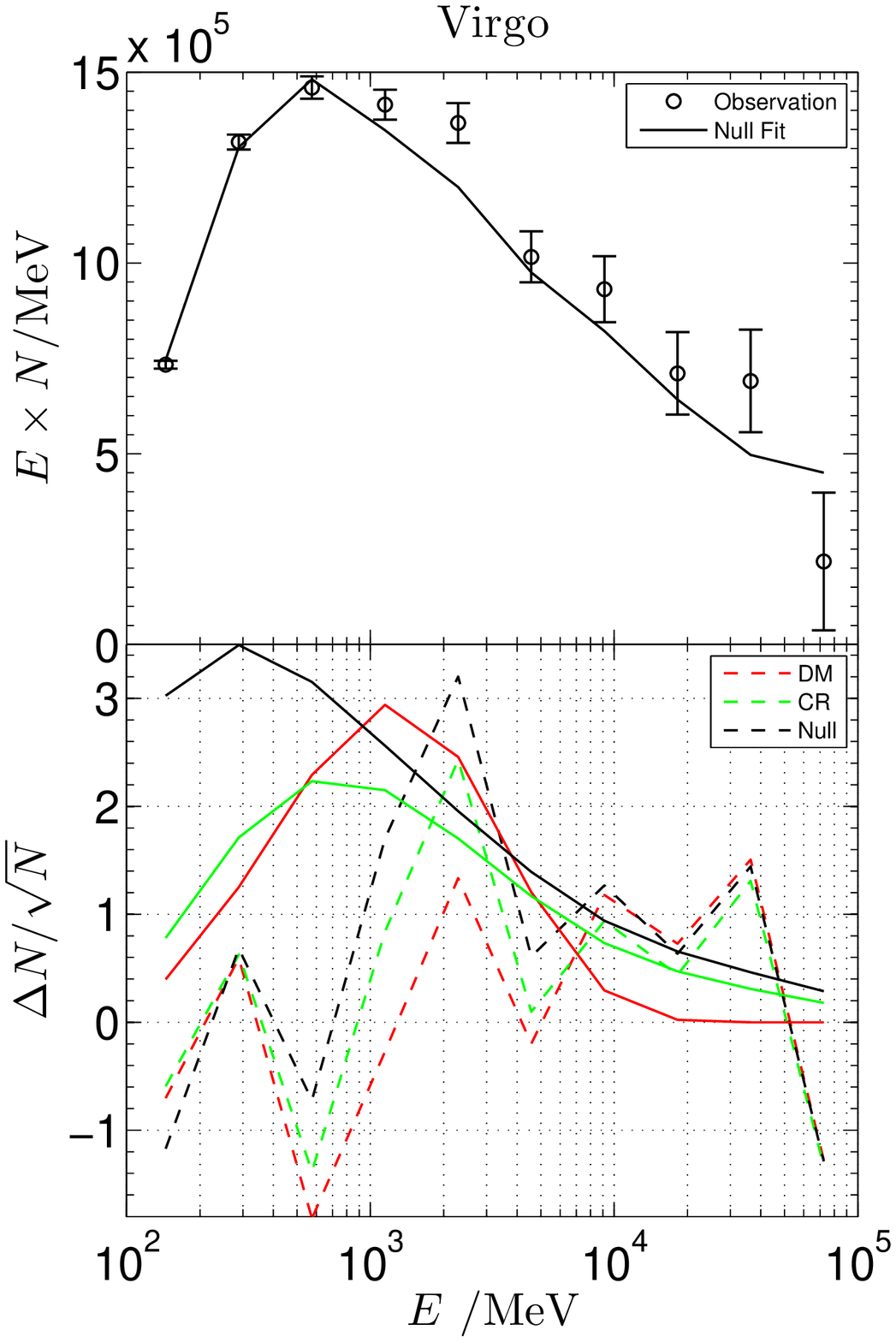}% 
\caption{Observed and fitted energy spectrum in the inner 3 degrees of
each cluster. The top panels show the counts (circles with error-bars)
and the null model (the model without CR or DM; solid line) fit in
each energy bin, multiplied by the energy of the bin to reduce the
dynamic range. The lower panels show the residual for the no-CR model
(black dashed line), the best-fit CR model without DM (green dashed)
and the best-fit DM model without CR (red dashed) for the case of $M_\chi\approx
30$~GeV annihilating into the \bbbar channel, normalized
by the estimated Poisson error in each bin. For comparison we also
show the contribution from the CR (green solid line) and DM (red solid
line) components in the corresponding models. For Virgo the black
solid line in the lower panel shows the contribution from the central
AGN in the null model. The best-fit parameters are taken from the
global best fit, i.e, from fitting the entire 10 degree region. Note
that we have rebinned the data into 10 energy bins to produce this
plot. }\label{f_SpecFit}
\end{figure}

\myfig
\includegraphics[width=0.33\textwidth]{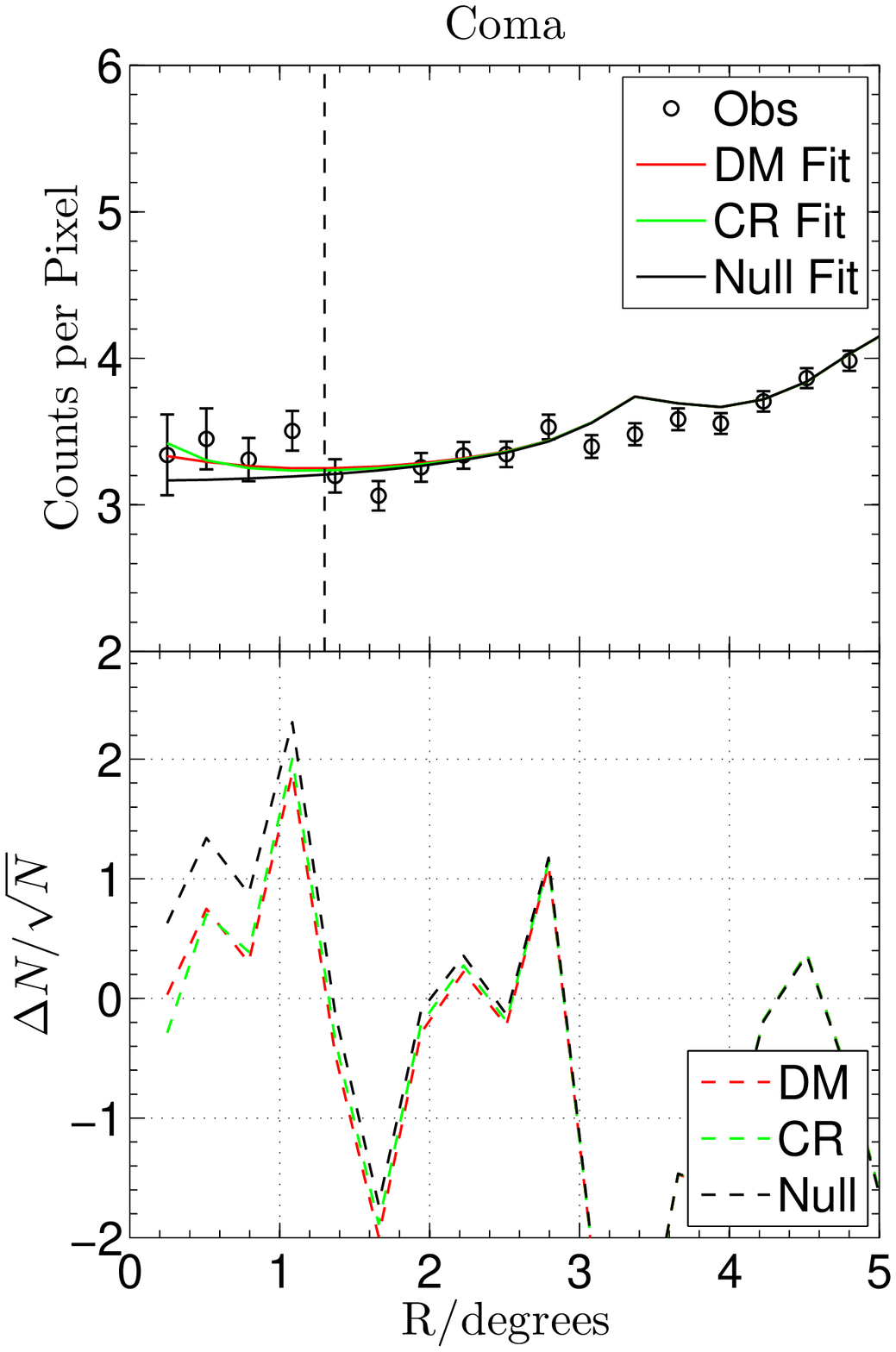}% 
\includegraphics[width=0.33\textwidth]{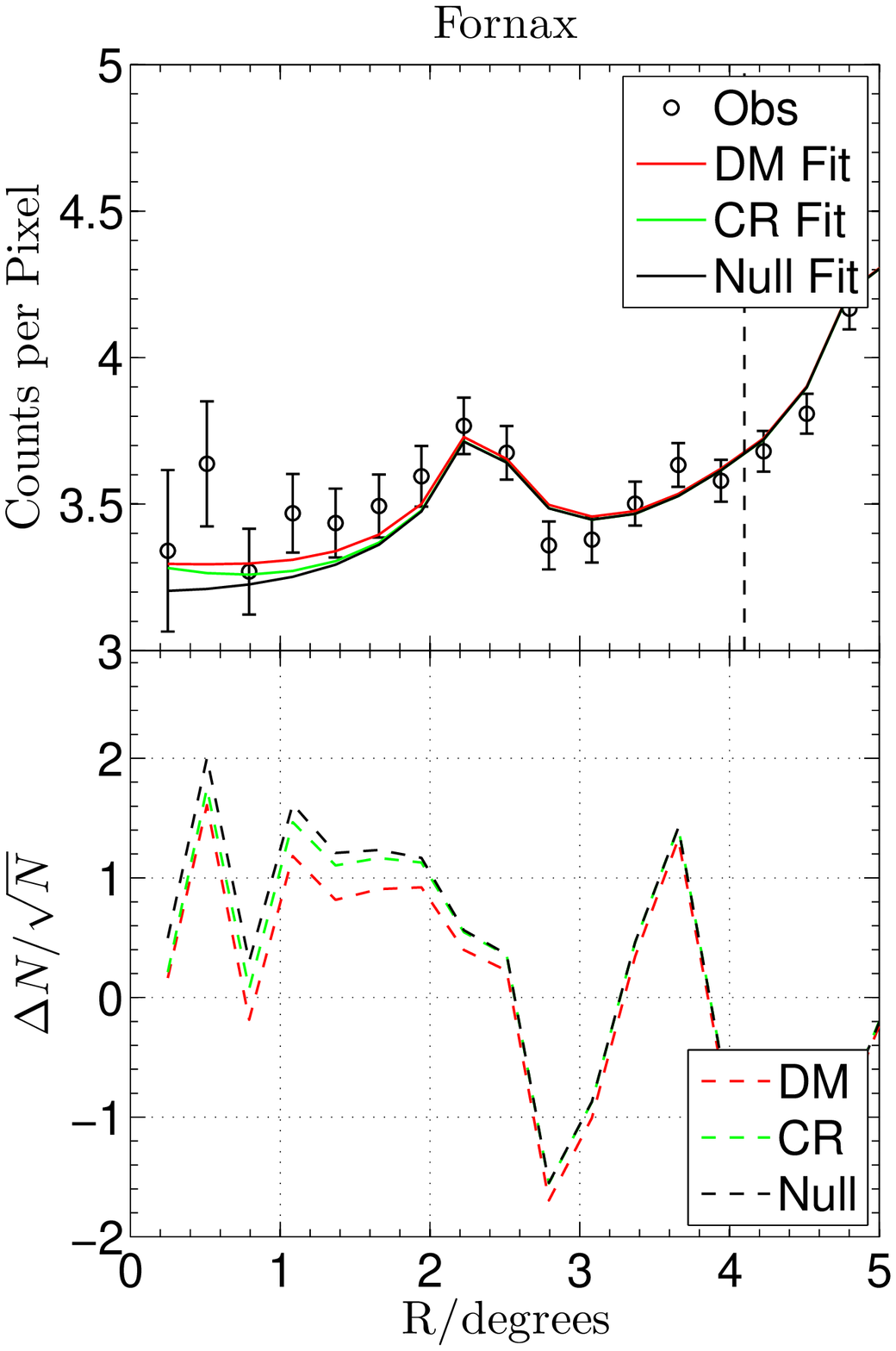}% 
\includegraphics[width=0.33\textwidth]{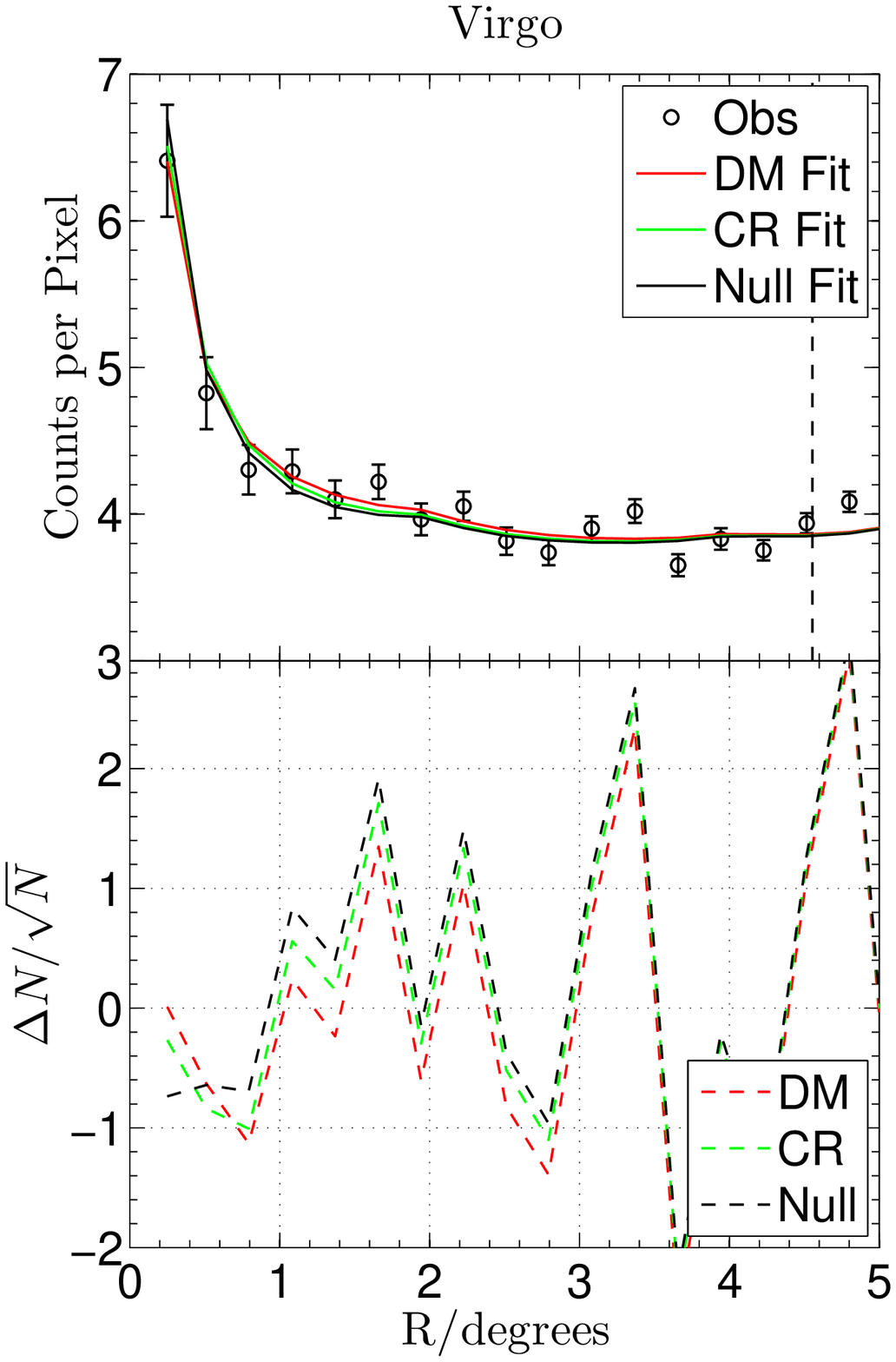}% 
\caption{Observed and fitted radial photon profile around each
cluster. The top panels show the counts (circles with error-bars) and
the fits for three models: the null model (i.e. without CR or DM;
black solid line), the DM-only model ($M_\chi\approx 30$~GeV
annihilating into the \bbbar channel, red solid line) and the CR-only
model (green solid line). The vertical dashed lines mark the cluster
virial radii. The lower panels show the residuals for the three
models, normalized by the estimated Poisson error in each bin. The
best-fit parameters are taken from the global best fit, i.e, from
fitting the entire 10 degree region.}\label{f_RadialFit}
\end{figure}

\myfig
\includegraphics[width=0.6\textwidth]{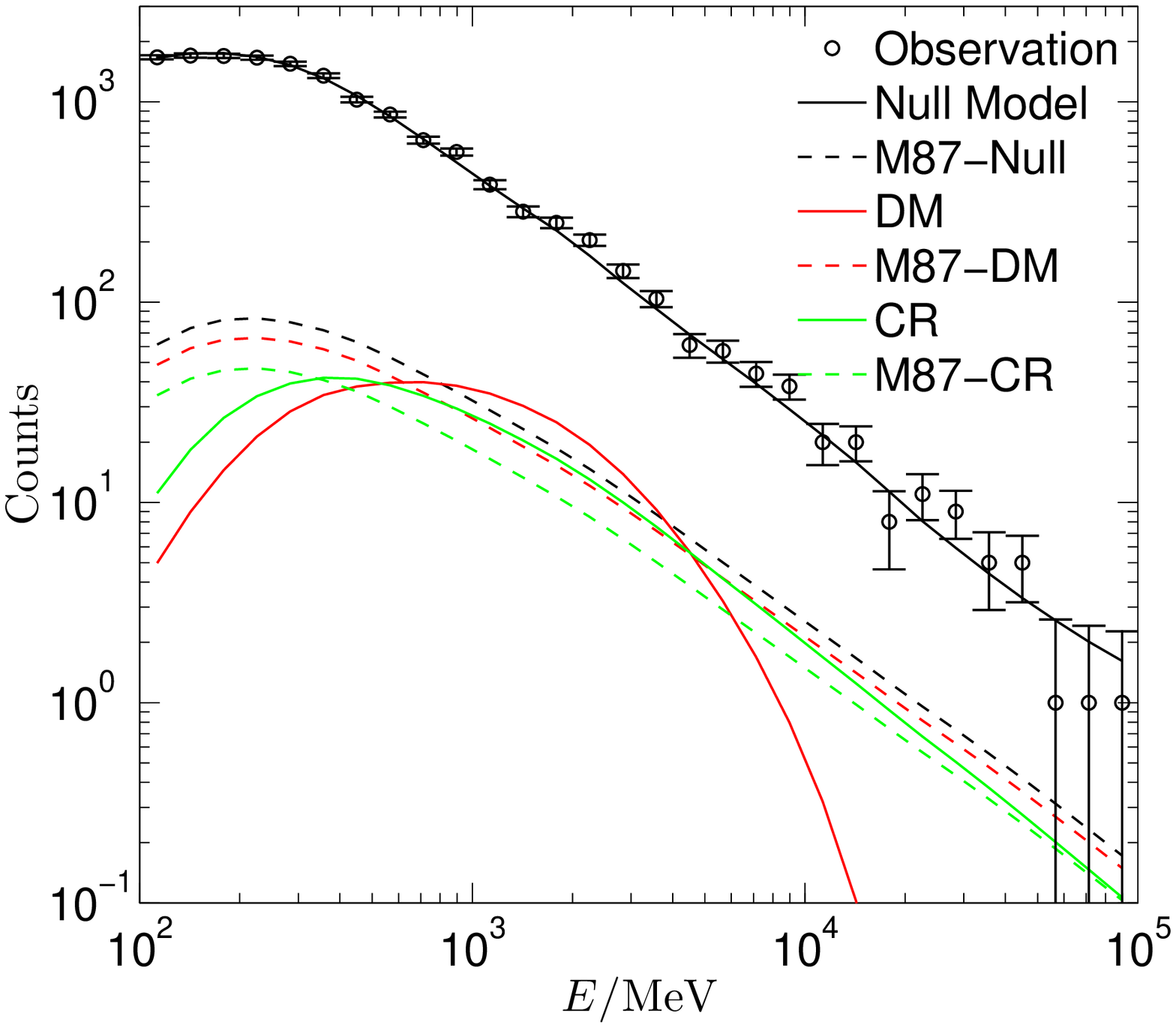} 
\caption{Contribution of the central AGN in Virgo (M87) in different
models. Circles with error-bars are the observed photon counts in each
energy bin. The black solid line shows the best-fit model without a DM
or CR component (the null model). The red solid line gives the counts
for the extended DM component in the DM-only model and the green solid
line gives the counts for the CR component in the CR-only model. The
three dashed lines give the fitted counts for M87 in the null (black),
DM-only (red) and CR-only (green) models respectively. Only counts
within a radius of three degrees are shown.}\label{fig_AGNspec}
\end{figure}
\section{Discussion and conclusions}

We have performed maximum likelihood fits to the 3-year Fermi-LAT data
for three galaxy clusters: Coma, Fornax and Virgo. We fit models
which, in addition to point sources and galactic and extragalactic
backgrounds, include emission due to dark matter (DM) annihilation and
cosmic rays (CR). For the former, we assume both a point source and
the theoretically predicted extended distribution of gamma rays in two
generic annihilation channels. When searching for a dark matter
signal, we experiment with different treatments of the CR
component. Our main results are as follows.

\begin{itemize}
\item Assuming no DM annihilation radiation, the gamma ray data
from Coma and Virgo already set significant constraints on the CR
level at more than 2$\sigma$. For Virgo, the data are consistent with
the predictions from the analytic CR model proposed by
\citet{Pinzke1,Pinzke2} while, for Coma, the data give an
upper limit that is a factor of two lower than the analytic
prediction, indicating either an uncertainty in model parameters such
as halo mass and gas density, or a peculiarity of the CR emission in
Coma. If interpreted as an error in the halo mass, a reduction in mass
by a factor of $1.6$ is required to reconcile the model with the upper
limits, assuming a simple CR luminosity scaling relation, $L_\gamma
\propto M_{vir}^{1.46}$ \citep{Pinzke1}, or a factor of $4.3$ according
to Equation~\ref{eq_CRR} in the case when the gas density profile is
fixed from X-ray observations. For Fornax, the non-detection of a CR
component is consistent with the low level predicted by the model.

\item Assuming no  CR emission, we obtain a detection of DM 
in Virgo at the $4.4\sigma$ significance level (TS=19.8) for a model
in which supersymmetric DM particles of mass $M_\chi\approx28$~GeV
annihilate into the \bbbar channel giving rise to an extended surface
brightness profile as predicted by the simulations of \cite{Gao}. For
Coma and Fornax the significance of a DM component is lower,
$2.3\sigma$ and $2.1\sigma$ respectively. The inclusion of cosmic rays
suppresses the likelihood of the DM annihilation component, depending
on the treatment of the CR emission. For Virgo the significance
decreases to $3.4\sigma$ when a free CR component is included and to
$2.9\sigma$ and $2.1\sigma$ respectively when our fiducial and optimal
CR models are included. The predicted DM annihilation flux and cross
section upper limits are more robust to the uncertainty in the CR
component for all three clusters, with the results for different CR
models agreeing to within a factor of two.

\item Assuming neither CR nor DM annihilation emission results in
excess emission within three degrees around each cluster. The spectrum
within this region shows that this excess peaks at around 1~GeV. These
two features combined lead to an increase in likelihood when an
extended emission component with an almost scale-free spectrum is
included in the model, explaining the significance seen in the CR and
DM components.

\item The DM model that best fits the spectral excess has a particle
mass in the range $20\sim60$~GeV for \bbbar final states or $2-10$~GeV
and $>1$~TeV for \mumu final states. These mass ranges are consistent
with the conclusions of the recent analysis of a gamma-ray excess in
the direction of the Galactic Center \citep{GalCen}. The preferred
masses are robust to changes in the treatment of cosmic rays. Within
these mass ranges, DM emission is preferred by the data over CR
emission in Virgo and Fornax.

\item Models in which the DM annihilation emission has the extended
profile predicted by cosmological simulations have a higher
significance and higher flux upper limits than models in which this
emission is assumed to be a point source. Due to the large boost
factor of the order of 1000, the cross-section upper limits for
extended models are at least 100 times lower than those for point
source models. Our cross-section constraints are much tighter than
those from an analysis of clusters using the 11-month data
\citep{FermiCluster} and are also tighter than those from a joint analysis
of the Milky Way's dwarf galaxies \citep{Dwarf,DwarfFermi}. Our new
limits exclude the thermal cross-section for $M_\chi<100$~GeV for
\bbbar final states and for $M_\chi<10$~GeV for \mumu final
states. The thermal cross-section, however, can still be reconciled
with the data by lowering the assumed boost factor from subhalos and
hence the $J$ factor. Since the boost factor, $b\propto
M_{cut}^{-0.226}$
\citep{Volker}, a cut-off mass of $10^{-4}\msun$, rather than our
assumed $10^{-6}\msun$, would be sufficient to
increase the cross-section limits by a factor of 3. 
\end{itemize} 

In our analysis we have chosen to adopt the values given in the LAT
2-year point source catalogue for the intensity and spectral shape of
the point sources lying within the virial radius of each cluster
(where the DM annihilation and CR emission levels are relatively
important). This allows for possible corrections to the point source
parameters in the presence of the CR and DM components, while also
avoiding the risk of re-fitting sources lying near the boundary of the
data region with less accuracy. However, we also tried keeping all the
point sources fixed or allowing the parameters of all the point
sources within the data region to vary during the fitting. We find
that this freedom in the treatment of the point sources has little
impact on the DM model fits. Our results are insensitive to the
uncertainty in cluster halo masses or concentration parameters. A
change in halo mass by a factor of 4, or a change in concentration
parameter by a factor of 2 would only change the overall normalization
of the inferred DM annihilation emission by a factor of 2 with the
profile shape remaining unchanged outside $0.01R_{vir}$. 
We also checked that the different energy cuts assumed in our analysis and in 
that of \citet{Huang} has no effect on the derived upper limits. We 
are able to reproduce the upper limits on the annihilation 
cross-section of \citet{Huang} for the test case of the Fornax
 cluster, after correcting for slightly different assumed boost factors.

The Virgo cluster where we have obtained our most significant DM
signal has a complex structure. Although we have concluded that the
central gamma-ray source in M87 does not interfere with our detection,
the existence of several neighboring structures (M86/M84, M49, M60,
M100), which show signs of interactions, could result in a
complex annihilation profile. Given the high significance of our
result for Virgo, it is worthwhile trying to improve the model for
this cluster as a multi-component structure and to use simulations of
merging clusters to improve the theoretical annihilation profile.

While we have shown that Virgo has an exceptionally high signal for
DM-like emission, the signal-to-noise ratio can be enhanced by
stacking several clusters. Such an analysis was recently carried
out by \citet{Huang}, but the signal-to-noise was degraded by the
assumption of an NFW annihilation profile rather than the extended
profile seen in the simulations. They considered an extended
subhalo-dominated annihilation profile but only for individual
clusters, not for the stack. Their stacked analysis resulted in looser
constraints on DM annihilation than their analysis of individual
clusters, presumably because the use of an inappropriate
theoretical profile resulted in the different clusters yielding
inconsistent results. Thus, it is clearly worth repeating the joint
analysis with the ``correct'' subhalo-dominated profile. It is also
tempting to extend the search for DM annihilation to multi-wavelength
data from the radio to very high energy gamma-rays.

The significance of the excess DM-type annihilation emission that we
have detected in our small cluster sample is only marginal and needs
to be confirmed with future observations. Nevertheless, the similarity
with the excess gamma-ray emission from the Galactic center and the
general trend for such excess emission in all three clusters examined
here is intriguing.

%outer profile modelling, cluster template size: the smaller template seems to give higher TS (2times) and higher cross-section and upperlimit (1.5times). but is it because the outer point sources are fixed(Answer: no!)? Another possibility is that the outer profile really drops faster than is modelled. The larger template has more feature, thus can be better constrained, perhaps revealing more discrepancy? try get a physical model for outer profile; effect of large scale structure.
 
%diffuse background model/2FGL pt model/Pass7 IRF cavets

%Virgo Relaxation: M86 1deg away, and M49 4deg away; Simulation profile applicable to this one?

%Halo parameter uncertainty (Mass/size/concentration/center)

\section*{Acknowledgments}
JXH is supported by the European Commissions Framework Programme
7,through the Marie Curie Initial Training Network Cosmo-Comp
(PITNGA-2009-238356). CSF acknowledges a Royal Society Wolfson
research merit award and an ERC Advanced Investigator grant. The
calculations for this work were performed on the ICC Cosmology
Machine, which is part of the DiRAC Facility jointly funded by STFC,
the Large Facilities Capital Fund of BIS, and Durham University. This
work was supported in part by an STFC rolling grant to the ICC. We
thank Shaun Cole, Jie Liu and Yu Gao for helpful discussions. JXH
acknowledges the support on software issues from Tesla Jeltema and the
Fermi science support team, especially Elizabeth C. Ferrara, Jeremy
S. Perkins and Dave Davis.

\bibliographystyle{\mybibstyle}
\bibliography{Fermi_ref}

\appendix
\section{Semi-Analytic formula for the Cosmic Ray induced gamma-ray emission}
Here we summarize the relevant equations for calculating the CR induced gamma-ray emission in galaxy clusters as derived in \citet{Pinzke1} and \citet{Pinzke2}. The CR induced photon source function from pion decay can be decomposed as: 
\begin{equation*}
\frac{dN_{\gamma}}{dtdVdE}=A(r)s(E). 
\end{equation*}
The spatial part is given by:
\begin{equation}\label{eq_CRR}
A(r)=((C_{vir}-C_{center})(1+(\frac{r}{R_{trans}})^{-\beta})^{-1}+C_{center}){\rho_{gas}(r)^2},
\end{equation}
with
\begin{eqnarray}
C_{vir} & = 1.7 \times 10^{-7} \times (M_{vir}/10^{15}M_\odot)^{0.51}\\
R_{trans} & = 0.021 R_{vir}\times (M_{vir}/10^{15}M_\odot)^{0.39}\\
\beta & =1.04 \times (M_{vir}/10^{15}M_\odot)^{0.15}
\end{eqnarray}
The spectrum is given as: 
\begin{eqnarray}
s(E) &
=g(\zeta_{p,max})D_{\gamma}(E_{\gamma},E_{\gamma,break})\frac{16}{3m_p^3c}\nonumber\\
     &\times \sum_{i=1}^3 \frac{\sigma_{pp,i}}{\alpha_i}
(\frac{m_p}{2m_{\pi^0}})^{\alpha_{i}}\Delta_i[(\frac{2E_\gamma}{m_{\pi^0}c^2})^{\delta_i}+(\frac{2E_\gamma}{m_{\pi^0}c^2})^{-\delta_i}]^{-\frac{\alpha_i}{\delta_i}},
\end{eqnarray}
with $\Delta= (0.767,0.143,0.0975)$, $\alpha=(2.55,2.3,2.15)$, $m_p$
the proton mass and
\begin{equation}
 D_\gamma(E_\gamma,E_{\gamma,break})=[1+(\frac{E_\gamma}{E_{\gamma,break}})^3]^{-1/9}. 
\end{equation}
The proton cut-off energy is 
\begin{equation}
E_{p,break}\approx \frac{10^8}{8} {\rm GeV} (\frac{R_{vir}}{1.5Mpc})^6, 
\end{equation}
and $P_{\gamma}\approx \frac{P_p}{8}$.
\begin{equation}
\sigma_{pp,i}\simeq 32(0.96+e^{4.42-2.4\alpha_i}) \mathrm{mbarn}. 
\end{equation}
The maximum shock acceleration efficiency is chosen to be $\zeta_{p,max}=0.5$ so that $g(\zeta_{p,max})=1$.
The gas density is fitted with multiple beta-profiles \citep{Pinzke2} as:
\begin{equation}
\rho_{gas}=\frac{m_p}{X_H X_e}\lbrace\sum_i
n_i^2(0)[1+(\frac{r}{r_{c,i}})^2]^{-3\beta_i}\rbrace^{1/2}, 
\end{equation}
where $X_H=0.76$ is the primordial hydrogen mass fraction and
$X_e=1.157$ is the ratio of electron and hydrogen number densities in
fully ionized ICM, with parameter values for $n_i(0)$, $r_{c,i}$ and
$\beta_i$ listed in TABLE VI of \citet{Pinzke2}.

\end{document}